\documentclass{comjnl}

\usepackage{url}
\usepackage{enumerate}
\usepackage{graphicx}
\usepackage{epstopdf}
\usepackage{multirow}
\usepackage{subfigure}
\usepackage{amsmath,amssymb}
\usepackage{booktabs}
\usepackage{multirow}
\usepackage{makecell}
\usepackage{caption2}

\usepackage{graphicx}  
\usepackage{float}  
\usepackage{subfigure}  

\begin{document}

\title[Association Rules Enhanced Knowledge Graph Attention Network]{Association Rules Enhanced Knowledge Graph Attention Network}
\author{Zhenghao Zhang}
\affiliation{School of Computer Science and Technology, Xidian University, Xi’an, 710071, China} \email{zhangzhenghao1108@126.com}

\author{Jianbin Huang\textsuperscript{*}}
\affiliation{School of Computer Science and Technology, Xidian University, Xi’an, 710071, China} \email{jbhuang@xidian.edu.cn}
\shortauthors{Jianbin Huang}

\author{Qinglin Tan}
\affiliation{School of Computer Science and Technology, Xidian University, Xi’an, 710071, China} \email{zhangzhenghao1108@126.com}



\begin{abstract}
Knowledge graphs enable a wide variety of applications, including information retrieval, question answering, and hypothesis generation. Despite the great effort invested in their creation and maintenance, most existing knowledge graphs suffer from incompleteness. Embedding knowledge graphs into continuous vector spaces has recently attracted increasing interest in knowledge base completion. However, in most existing embedding methods, only fact triplets are utilized, and logical rules have not been thoroughly studied for the knowledge base completion task. To overcome the problem, we propose an association rules enhanced knowledge graph attention network (AR-KGAT). The AR-KGAT captures both entity and relation features for high-order neighborhoods of any given entity in an end-to-end manner under the graph attention network framework. The major component of AR-KGAT is an encoder of an effective neighborhood aggregator, which addresses the problems by aggregating neighbors with both association-rules-based and graph-based attention weights. Additionally, the proposed model also encapsulates the representations from multi-hop neighbors of nodes to refine their embeddings. The decoder enables AR-KGAT to be translational between entities and relations while keeping the superior link prediction performance. A logic-like inference pattern is utilized as constraints for knowledge graph embedding. Then, the global loss is minimized over both atomic and complex formulas to achieve the embedding task. In this manner, we learn embeddings compatible with triplets and rules, which are certainly more predictive for knowledge acquisition and inference. We conduct extensive experiments on two benchmark datasets: WN18RR and FB15k-237, for two knowledge graph completion tasks: the link prediction and triplet classification to evaluate the proposed AR-KGAT model. The results show that the proposed AR-KGAT model achieves significant and consistent improvements over state-of-the-art methods.
\end{abstract}

\maketitle

\section{Introduction}
\label{section1}
Over recent years, large-scale knowledge base, such as Freebase \cite{1}, DBpedia \cite{2}, and YAGO, have been developed to store structured information of common knowledge. KBs are represented as directed multi-relational graphs, Knowledge Graphs (KGs), where entities and relations are represented as nodes and as edges of different types, respectively. They usually consist of numerous facts structured in the triplets: $<$head entity, relation, tail entity$>$, e.g.  $<$Melbourne, city-of, Australia$>$ and $<$Paris, Capital-of, France$>$.

Through rich factual knowledge to be organized and stored, KGs play a critical role in various natural language processing (NLP) applications in recent years, including question answering (\cite{3}, \cite{7}), machine reading \cite{4}, information retrieval, and semantic searching (\cite{5}, \cite{6}), and personalized recommendation \cite{7}. Many large-scale KGs have been constructed with millions of entities and relations. However, the existing facts and newly created knowledge are too vast, and thus a lot of valid triplets are missed in the real world \cite{8}. For example, nationalities or birthplaces are missed in more than 70\% of the person entries in Freebase. As shown in Figure 1, the nationality of Kobe Bryant (a famous basketball star) is missed. Thus, the target triplet (Kobe Bryant, nationality, ?) does not exist as a knowledge graph to be retrieved.

KGs are inherently discrete and incomplete. To alleviate this drawback, relation prediction, also referred to as knowledge base completion, infers the missing facts based on the given facts. Enormous efforts have been devoted to knowledge base completion to estimate missing relations between entities under the supervision of the existing knowledge graph. 

Knowledge graph embedding (KGE) successfully handled the problems of symbolic nature in various KGs. In KGE, the components of KG: entities and relations were embedded into a low dimensional continuous vector space, while specific properties of the original graph were preserved. Accordingly, the inherent structure of the KG is preserved, while the manipulation is simplified. KGE has recently attracted increasing interest in knowledge base completion and inference, with progressive advancement from the translational models (TransE\cite{10}, TransH\cite{11}, and DistMult\cite{13}) to the recent deep CNN models (e.g., ConvE\cite{18} and ConvKB\cite{33}). However, in these embedding models, each triplet was independently processed, resulting in the loss of the potential information of knowledge base. Rich semantical and latent relationships between them have not been exploited, which are inherently implicit in the local neighborhood surrounding a triplet. 

Recently, the Graph Neural Network (GNN) was proposed to utilize the graph connectivity structure as another way of learning graph node embedding. However, in most of the existing researches on GNNs, node representations were learned in simple undirected graphs. Since KGs are multi-relationally structured, unlike homogeneous graphs\cite{9}, naïve application of the existing GNN models in handling relational graphs induces over-parameterization. Also, it is limited to only node representations learning. Therefore, it is required to advance the GNN framework to jointly learn node and relation representations by utilizing KG embedding techniques. 

\begin{figure}[H]
	\centering 
	\includegraphics[width=0.75\columnwidth]{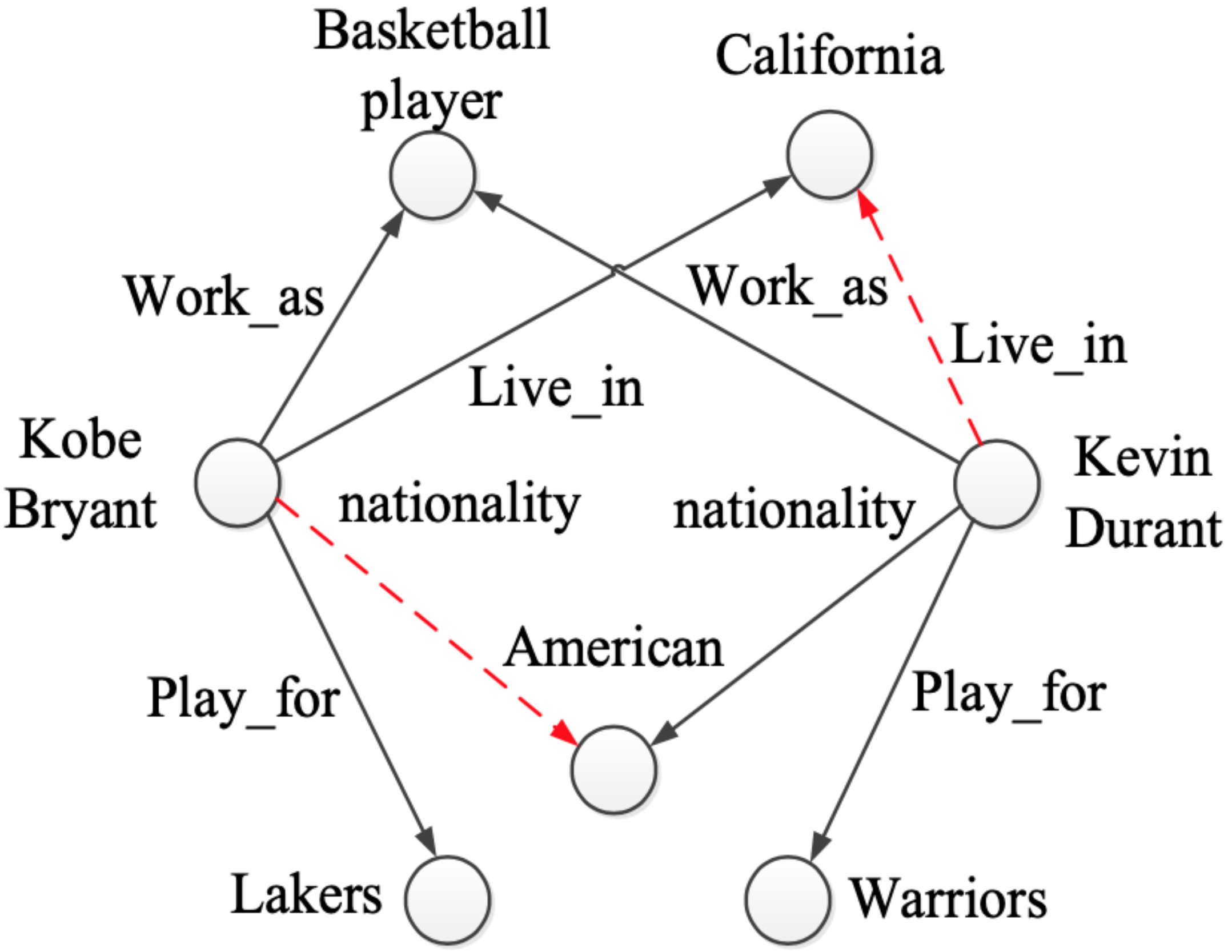}
	\caption{Example of a knowledge graph. Sub-graph consists of the relationships between entities (solid lines) and inferred relationships (dashed lines). Inferred relationships are initially hidden.}
	\label{fig_1}
\end{figure}

Moreover, many existing methods focused only on fact-triplets from the given knowledge graph. The models have the following limitations: (1) the logical rules obtained from KB do not lead to other scientific basis or the reduced sparseness of knowledge graphs. (2) joint embedding of the rules and knowledge graph is challenging since the traditional algebraic operations and rules of logic symbols cannot be naively used in triplets. A flexible and declarative language is provided by the logical rules for expressing rich background knowledge. Thus, to transfer human knowledge to entity and relation embedding, the logical rules can be integrated into knowledge graph embedding, strengthening the learning process. 

In order to overcome the limitations, we propose an end-to-end GAT framework for multi-relational knowledge graphs, Association Rules Enhanced Knowledge Graph Attention Network (AR-KGAT). The proposed AR-KGAT generalizes and extends the attention mechanism by taking the benefits of both logic-based and graph-based attention simultaneously. Also, entity-relation composition operations are systematically leveraged from knowledge graph embedding techniques for relation prediction. Specifically, the proposed AR-KGAT framework consists of three main designs to correspondingly address the challenges above in knowledge graph embedding. \textbf{First}, our uniform framework considers two types of association rules: one-to-one and n-to-one rules to fully utilize rich information of rules to augment knowledge inference. Then, through the mining algorithm, only rules with confidences greater than a threshold are selected in our embedding model. As a result, the sparseness of knowledge graphs is reduced. \textbf{Second}, in the proposed novel aggregator, the neural attention mechanism and association rules-based weighting mechanism are simultaneously used to learn the weights of the neighborhood of any given entity. Also, we propose the recursive embedding propagation to update an entity’s embedding based on its neighborhood. Through the recursive embedding propagation, high-order relations are captured in linear time complexity. \textbf{Last}, the AR-KGAT is learned by minimizing a global loss consisting of triplets and rules terms to obtain entity and relation embeddings. The learned embeddings can better predict knowledge acquisition and inference since it is compatible with both triplets and logical rules. 

Extensive comparisons are conducted to evaluate the proposed rule-enhanced method under two benchmark datasets on several KG completion tasks: link prediction and triplet classification. The experimental analysis shows that our model provides remarkable performance gains compared to state-of-the-art methods. 

The contributions of this paper are as follows:

(1)Triplets and logical rules are jointly modeled in the proposed unified framework to achieve more predictive entity and relation embeddings. 

(2) Two types of association rules are introduced, and the association rules and the corresponding promotion degree are automatically mined. 

(3) KG-specific attention based neighborhood aggregator is proposed. It recursively propagates representations along with high-order connectivity from an entity’s neighbors to update its representation. Weights of cascaded propagations are generated through the integration of logic- and graph-based attention mechanisms to reveal the importance of such high-order connectivity. 

(4) The representations for entity and relation are learned by a global loss function compatible with both triplets and high-order logical rules. 

The remainder of this paper is organized as follows. Section 2 briefly reviews the related works. Section 3 presents the preliminaries of our proposed model. The details of our approach are described in Sections 4 and 5. The experimental results of the proposed model are analyzed in section 6. Section 7 concludes this paper with a future research direction.

\section{Related work}
\label{sec:related-works}
Recently, several variants of knowledge graph embedding methods have been proposed for relation prediction. These methods can be broadly classified as: translational based, convolutional neural network based, and graph neural network based models. Moreover, existing research of jointly embedding KG and logical rules are also introduced.

\subsection{Translational Models}\label{subsecHINModel}
Starting with TransE, there have been multiple proposed approaches that use simple operations like dot products and matrix multiplications to compute a score function. Other transition-based models extend TransE to additionally use projection vectors or matrices to translate head and tail embeddings into the relation vector space, such as: TransH \cite{10}, TransR \cite{11}, and TransD \cite{12}, STransE \cite{31}. TransH allows entities to have multiple representations. To obtain multiple representations of an entity, TransH projects an entity vector into relation-specific hyperplanes. TransR also handles the problem of TransE by introducing relation spaces. It allows an entity to have various vector representations by mapping an entity vector into relation-specific spaces. To handle multiple types of relations, TransD constructs relation mapping matrices dynamically by considering entities and a relation simultaneously. However, these models lose the simplicity and efficiency of TransE. Furthermore, DISTMULT \cite{13} and ComplEx \cite{14} use a tri-linear dot product to compute the score for each triple. Recent research has shown that using relation paths between entities in the KBs could help to get contextual information for improving KB completion performance (\cite{6}, \cite{15,16,17}). These translational models are faster, require fewer parameters and are relatively easier to train, but result in less expressive KG embeddings. Translational approaches on KG embedding work in a transductive manner. They require that all entities should be seen during training. Such limitation hinders them from efficiently generalizing to emerging entities.

\subsection{Convolution Based Models}\label{subsecHINModel}

Translation based embedding models are a popular form of representation model. While translational models learn representations using simple operations and limited parameters, they produce low quality representations. Shortcomings of translation based models however, limits their practicability as knowledge completion algorithm. In contrast, Convolution based models learn more expressive representations due to their parameter efficiency and consideration of complex relations. Several recent works suggest that convolutional neural network (CNN) based models generate richer and more expressive feature embeddings and hence also perform well on relation prediction. Dettmers et al. \cite{18} proposed ConvE—the first model applying CNN for the KB completion task. ConvE uses stacked 2D convolutional filters on reshaping of entity and relation representations, thus increasing their expressive power, while remaining parameter efficient at the same time. ConvKB \cite{33} is another convolution based method which applies convolutional filters of width l on the stacked subject, relation and object embeddings for computing score. 

However, both translational and CNN based models still have not fully exploited the potential of the knowledge base since they suffer from the following limitations: these methods process each triple independently and hence fail to encapsulate the semantically rich and latent relations that are inherently implicit in the local neighborhood surrounding a triple. 

\subsection{Graph Neural Network Based Models}
In \cite{19}, the authors propose applying Graph Neural Networks (GNNs) on the KG, which generates the embedding of a new entity by aggregating all its known neighbors. In addition, two popular variants of GNN, including Graph Convolutional Networks (GCN) and Graph Attention Networks (GAT), have been proposed to address this shortcoming in modeling graph-structured data which aggregate local neighborhood information for each node, and successfully applied to various domains. However, their model aggregates the neighbors via simple pooling functions, which neglects the difference among the neighbors. Other works like \cite{22} aim at embedding nodes for node classification given the entire graph and thus are inapplicable for inductive KG-specific tasks. A graph based neural network model called R-GCN \cite{20} is an extension of applying graph convolutional networks (GCNs) \cite{21} to relational data. It applies a convolution operation to the neighborhood of each entity and assigns them equal weights. This graph based model does not outperform the CNN based models. Existing methods either learn KG embeddings by solely focusing on entity features or by taking into account the features of entities and relations in a disjoint manner. 

\subsection{Jointly embedding KG and logic rules}
Most existing methods embed the knowledge graph based solely on triples contained in the knowledge graph. Several recent works try to incorporate other types of available knowledge, e.g., relation paths (\cite{23}, \cite{15}), relation type-constraints \cite{24}, entity types \cite{12}, and entity descriptions \cite{25}, to learn better embeddings. Logic rules have been widely used in knowledge inference and acquisition \cite{40,41}, usually on the basis of Markov logic networks (\cite{26,27,28}). Recent works have shown that the inclusion of background knowledge, such as logical rules, can improve the performance of embeddings in downstream machine learning tasks. There has been growing interest in combining logical rules and embedding models. Wei et al. \cite{30} tried to leverage both embedding methods and logical rules into the knowledge graph embedding for KG completion. In their work, however, rules are modeled separately from embedding methods, serving as post-processing steps, and thus will not help to obtain better embeddings. Rocktäschel et al. \cite{31} proposed a joint model which jointly encodes the rules into the embedding. In \cite{32} a new method named LR-KGE is proposed to jointly embed the knowledge graph and logic rules. However, their work focuses on relation extraction task and creates embeddings for entity pairs, and hence fails to discover relations between unpaired entities.  
                  
\section{Preliminaries}\label{sec:proposed-model}
We begin this section by introducing the notations and definitions of Knowledge Graph (KG) used in the rest of the paper, followed by a brief overview of Graph Attention Network (GATs) for undirected graphs and its extension to knowledge graphs. 

\subsection{Knowledge Graph}
Knowledge Graphs (KGs) can be represented by a collection of valid factual triples in the form of (\textit{head entity, relation, tail entity}) denoted as $(h, r, t) .$ Each triple consists of two entities $h, t \in E$ and a relation $r \in R, h$ denotes the head entity of a triple and $t$ denotes the tail entity of a triple.
Moreover, for each entity $e,$ we denote by $N_{K(e)}$ its neighborhood in $K,$ i.e., all related entities with the
involved relations. Formally, $N_{K(e)}=\left\{\left(r, e^{\prime}\right) \mid\left(e, r, e^{\prime}\right) \in K\right\} .$ We denote the projection of $N_{K(e)}$ on
$E$ and $R$ by $N_{E(e)}$ and $N_{R(e)},$ respectively. Here $N_{E(e)}$ are neighbors and $N_{R(e)}$ are neighboring
relations.

\subsection{Knowledge Graph Embedding}
Knowledge graph embedding is an effective way to parameterize entities and relations as vector  representations, while preserving the graph structure. Embedding models try to learn an effective low-dimensional representations of entities, relations, and a scoring function $f,$ such that for a given input triple $(h, r, t), f$ gives the likelihood of $(h, r, t)$ being a valid triple by optimizing the translation principle $e_{h}^{r}+e_{r} \approx e_{t}^{r} .$ Finally, to learn the entity and relation representations, an optimization problem is solved for maximizing the plausibility of the triple in the KG.

Let us take a widely used knowledge graph embedding model TransE (Bordes et al., 2013) as an  example to illustrate it explicitly: TransE employs a transitional characteristic to model relationships between entities, in which it assumes that if $(h, r, t)$ is a valid fact, the embedding of head entity $h$ plus the embedding of relation $r$ should be close to the embedding of tail entity $t,$ i.e. $g(h, r, t)=\left\|v_{h}+v_{r}-v_{t}\right\|_{p}^{p}$ of the positive triple $(h, r, t)$ should be close to 0 and smaller than score of negative triples $\left(h^{\prime}, r, t^{\prime}\right) .$ Herein, $v_{h}, v_{t} \in \mathbb{R}^{d}$ and $v_{r} \in \mathbb{R}^{k}$ are the embedding for $h, t$ and $r$ respectively. Moreover, the negative triple $\left(h^{\prime}, r, t^{\prime}\right)$ is obtained from $(h, r, t)$ by replacing $h$ by $h$ ' or $t$ by $t^{\prime}$. A lower score of $g(h, r, t)$ suggests that the triplet is more likely to be valid, and vice versa. The training of TransE aims to optimize the discrimination between positive triplets and negative ones.

\subsection{Graph Attention Networks}
Graph convolutional networks gather information from the entity’s neighborhood and all neighbors  contribute equally in the information passing. To address the shortcomings of GCNs, introduced Graph Attention Networks (GATs). GATs learn to assign changing levels of importance to nodes in every  node’s neighborhood, rather than treating all neighboring nodes with equal importance, as is done in GCN. The input feature set of nodes to a layer is $x=\left\{\vec{x}_{1}, \vec{x}_{2}, \ldots, \vec{x}_{N}\right\}$. A layer produces a transformed set of node feature vectors $x^{\prime}=\left\{\vec{x}_{1}^{\prime}, \vec{x}_{2}^{\prime}, \ldots, \vec{x}_{N}^{\prime}\right\},$ where $\vec{x}_{i}$ and $\vec{x}_{i}^{\prime}$ are input and output
embeddings of the entity ei, and $\mathrm{N}$ is number of entities, a single GAT can be described as:
$$
e_{i j}=a\left(W \vec{x}_{i}, W \vec{x}_{j}\right),
$$
where $e_{i j}$ is the attention value of the edge in $G, W$ is a parameterized linear transformation matrix mapping the input features to a higher dimensional output feature space, and $a$ is any attention function  of our choosing. Attention values for each edge are the importance of the edge’s features for a source 
node. Here, the relative attention value can be computed using a softmax function over all the values in the neighborhood.

\section{Association Rules Mining}\label{sec:Multi-View Dynamic HIN}

\begin{figure*}[!t] 
	\setlength{\abovecaptionskip}{0cm}
	\setlength{\belowcaptionskip}{-0.2cm}
	\centering 
	\includegraphics[width=0.85\linewidth]{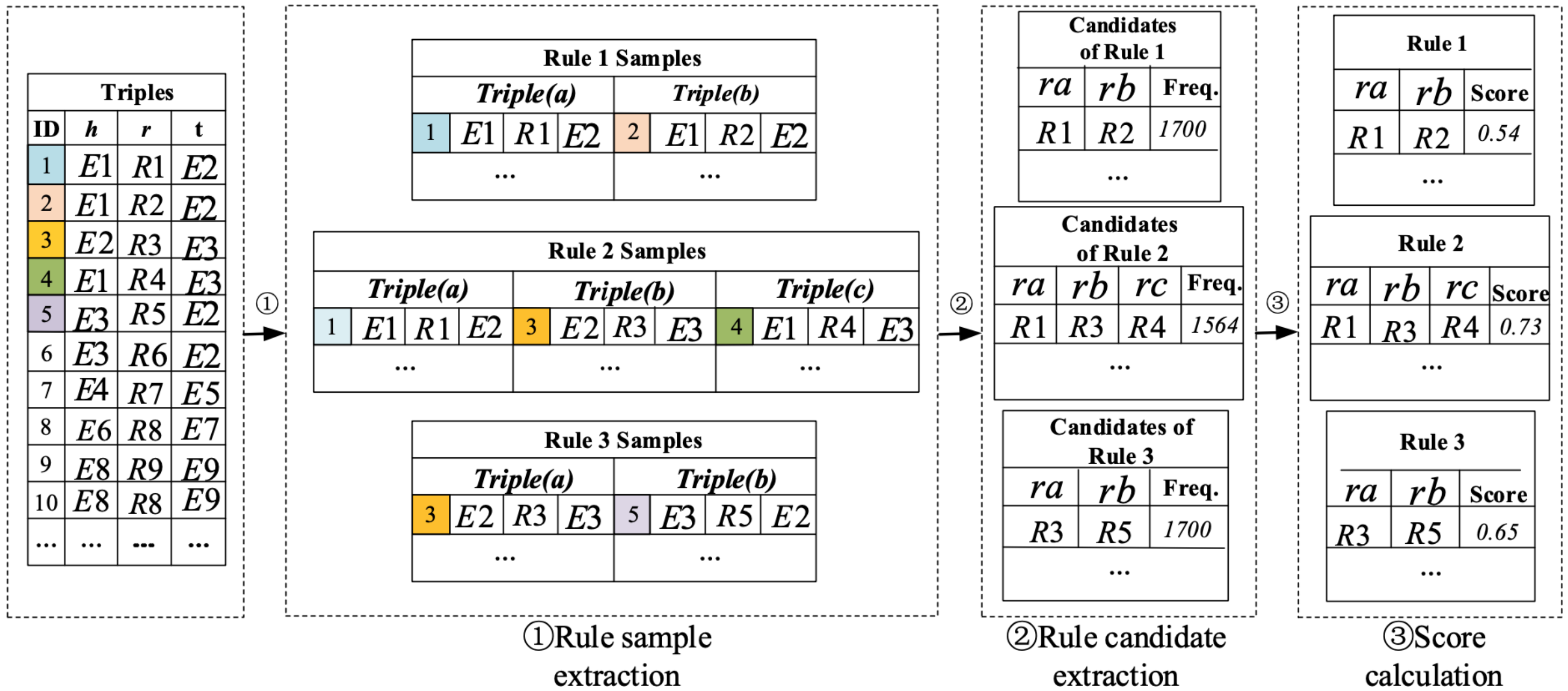} 
	\caption{The overall framework of association rules mining from a knowledge graph.} 
	\label{arc} 
\end{figure*}

In the proposed framework, the input and output are triplets of the knowledge graph $G$ and the association rules of the two types with corresponding scores. The main steps of the proposed framework 
are described in this section as follows.

\subsection{Rules Modeling}
A knowledge base $(\mathrm{KB})$ is represented by a set of triples $T=\{(h, r, t)\}$ with each triple consists of two entities $h, t \in E$ and a relation $r \in R . E$ stands for the set of entities while $\mathrm{R}$ stands for the set of relations. For any triple $\left(e_{i}, r_{k}, e_{j}\right)$ is taken as a ground atom which applies a relation $r_{k}$ to a pair of entities $\left(e_{i}, e_{j}\right)$ and can be modeled by the translation assumption, i.e., relations act as translations between head and tail entities. Moreover, triples are represented as atomic formulae and modeled by the translation assumption, while $n$-hop logical rules can then be interpreted as a complex formula $f$, constructed by combining a set of ground atoms(triples) with logical connectives, and modeled by t-norm fuzzy logics.

A rule can present an abstract pattern mined from the data. It is found that rules in a KG are simply not independent of each other. For an entity $e,$ one neighboring relation $r_{1}$ may imply the existence of another neighboring relation $r_{2}$. Taken Figure \ref{fig_1} as example, the fact that one ``plays for'' Los-Angeles
Lakers always implies that ``he is a Basketball Player of Lakers''. Additionally, a neighbor relation ``play for'' Los Angeles could help to imply the relation of "live in" since there are other athletes playing for Lakers while living in Los-Angeles city. Hence, it is significantly beneficial to exploit such association
rules in an entity's neighborhood to make the aggregation more informative.

In this paper, we consider two types of association rules, which will be used in our proposed model:

\textbf{\textit{Definition 1. (One-to-One Association Rules):}} One-to-One Association rules can be divided into two categories:

\textbf{\textit{Type 1 (Inference Rule):}} inference rule discovers correlation of one-hop relations with the same head and tail entities. This Association rule can be represented in the form of $\forall x, y:\left(x, r_{s}, y\right) \Rightarrow\left(x, r_{t}, y\right)$, stating that any two entities linked by relation $r_{s}$ should also be linked by relation $r_{t}$.

For instance, a universally quantified rule $\forall x, y:(x, capital-o f, y) \Rightarrow(x, located-i n, y)$ might be instantiated with the concrete entities of Beijing and China, giving the ground rule:
$$
(Beijing,Capital-of,China)  \Rightarrow  (Beijing,located-in,China) 
$$

\textbf{\textit{Type 2 (Anti-symmetry Rule):}} An anti-symmetry rule can be denoted in the form of $\forall x, y:\left(x, r_{a}, y\right) \Rightarrow\left(y, r_{b}, x\right) .$ The relation $r_{a}$ can imply the anti-symmetry relation $r_{b},$ which denotes
that two relations $r_{a}$ and $r_{b}$ are anti-symmetrical. For example, (apple, hypernym, fruit) $\Leftrightarrow$ (fruit, hyponym, apple). Specifically, an anti-symmetry rule is undirected.

\textbf{\textit{Definition 2. (N-to-One Association Rules):}} N-to-One association rule is quite similar in nature 
from one-to-one association rules we introduce above. When n=2, a specific two-to-one transitivity rule can be represented by:
$$
\forall x, y, z:\left(x, r_{s_{1}}, y\right) \wedge\left(x, r_{s_{2}}, z\right) \Rightarrow\left(x, r_{t}, z\right).
$$
This transitivity rule above denotes that if $x$ and $y$ are linked by relation $s_{1}$ and $y$ and $z$ are linked by $s_{2}, x$ and $z$ will be linked by relation $t$.

For instance, a KB may contain the fact that a child has a mother, then the mother’s husband is most likely the father. In Figure \ref{fig_1}, we define an instance for the entities Kobe, Vanessa and Natalia, and they are connected through the triples including (Vanessa, is mother, Natalia), (Kobe, is father, Natalia), (Kobe, marry with, Natalia). In the example, we can mine the two-to-one transitivity rule: $\left(x_{1}, hasamother , x_{2}\right) \wedge\left(x_{2},  marrywith , x_{3}\right) \Rightarrow\left(x_{1},  hasafather,  x_{3}\right)$

It is well known that rules are important for reasoning and they are also useful for KG completion, 
which is the process to automatically extract new facts from existing ones (e.g., link prediction). In this 
paper, both one-to-one and n-to-one association rules are used to discover correlations in Knowledge 
Base (KB). The goal of our paper is to incorporate such two types of association rules from KBs in 
graph neural network framework and evaluate whether these rules can indeed improve the quality of 
embedding. Moreover, it is easy to see that besides these two types of rules, our framework is general 
enough to handle other types of rules. The investigation of other types of association rules will be left for 
future work.

\subsection{Rules Extraction}
LR-KGE \cite{32} was proposed to manually filter the top-ranked wrong rules since the results inferred 
by wrong candidate rules are not reasonable. However, LR-KGE cannot be applied to a large-scale 
knowledge graph having a large number of relations. In this paper, a novel approach is proposed to choose proper rules from the candidate pool.

\subsubsection{Extraction of Rule Sample}
Samples of the rule are extracted from given triplets in this step. The sample rule is referred to as the triplet combinations that satisfy specific conditions. For example, a one-to-one rule sample comprises the triplet $t_{1}$ and triplet $t_{2}$ in the 
``One-to-One Rule Samples'' table of Figure $2$. The rule sample  satisfies the one-to-one association rule that the two triplets have the same head and tail entities. On the other hand, the n-to-one rule sample comprises the triplet $t_{1}$, triplet $t_{3}$, and triplet $t_{4}$ in the N-to-One Rule Samples table of Figure \ref{fig_2}. The rule sample satisfies the n-to-one association rule that the entities are sequential. Through this method, the rule samples that belong to the above rule types can be mined.

\subsubsection{Extraction of Rule Candidate}
Statistics are derived from the extracted rule candidates. For instance, we extract $\left(r_{1}\right.$ and $\left.r_{2}\right)$ in the ``Candidates of One-to-One Rule'' from each one-to-one association rule sample. Also, $\left(r_{1}, r_{3}\right.$ and
$r_{4}$ ) in the ``Candidates of N-to-One Rule'' are extracted from each n-to-one association rule sample. Those candidates for one-to-one and n-to-one association rules are extracted in such a way. Then, the candidates are ranked in terms of the frequencies in descending order.

\subsection{Calculation of Promotion Degrees}
Here, $f_{a}$ and $f_{b}$ are defined as two constituent formulas, either atomic or complex, which consist of a single atom or a set of atoms (triples) with logical connectives. Following notations in logics, we denote potential dependency between $f_{a}$ and $f_{b}$ by an ``association rule'' $\left(f_{a} \Rightarrow f_{b}\right)$. To measure the extent of such dependency, we will define three metrics, which are denoted as $p_{support}\left(f_{a} \Rightarrow f_{b}\right),  p_{confidence}\left(f_{a} \Rightarrow f_{b}\right)$ and $p_{promotion}\left(f_{a} \Rightarrow f_{b}\right)$ of $\left(f_{a} \Rightarrow f_{b}\right)$ respectively. And $p\left(f_{a}\right),  p\left(f_{b}\right)$ are proportions of entities with formulae $f_{a}$ and $f_{b}$ to all entities respectively.

\textbf{\textit{Definition 3 (Support Degree):}} In a KG, $p_{support}\left(f_{a} \Rightarrow f_{b}\right)$ indicates the proportion of association rule $\left(f_{a} \Rightarrow f_{b}\right),$ which denotes that entities with formula $f_{a}$ also have $f_{b}$ as a neighboring formula to all entities, and can be denoted as follows:
$$
p_{support}\left(f_{a} \Rightarrow f_{b}\right)=\dfrac{n_{e}\left(f_{a} \wedge f_{b}\right)}{N}.
$$
where $N$ denotes as the total number of entities, and $n_{e}\left(f_{a} \wedge f_{b}\right)$ represents the number of entities with
neighboring formula $f_{a}$ also have $f_{b}$. As an empirical statistic over the entire $\mathrm{KG}, p_{\text {support}}\left(f_{a} \Rightarrow f_{b}\right)$ is larger if more entities with neighboring formula $f_{a}$ also have $f_{b}$ as an association rule. For example, there are 10000 entities in a $\mathrm{KB}$ Dataset, including 6000 entities with neighbor formula $f_{1}$,
7500 with $f_{2},$ and 4000 with both. Then we can obtain the support degree:
$$
p_{support}\left(f_{1} \Rightarrow f_{2}\right)=\dfrac{4000}{10000}=0.4.
$$

\textbf{\textit{Definition 4 (Confidence Degree):}} It indicates the proportion of entities with neighboring formula both $f_{a}$ and $f_{b}$ at the same time to entities that have $f_{a}$ as a neighboring formula, and can be denoted as follows:
$$
p_{Confidence}\left(f_{a} \Rightarrow f_{b}\right)=\dfrac{n_{e}\left(f_{a} \wedge f_{b}\right)}{n_{e}\left(f_{a}\right)}.
$$

We continue to take the previous example as illustration, $p_{Confidence}\left(f_{1} \Rightarrow f_{2}\right)=\dfrac{4000}{6000}=0.67,$ which denotes that $67 \%$ of entities with $f_{1}$ have $f_{2}$ as a neighboring formula. As we can see above example,
the confidence degree of $f_{1}$ to $f_{2}$ is $0.67,$ which seems to be quite high, but in fact, it is misleading. Why do you say that? Because in the absence of any conditions, the proportion of $f_{2}$ is $0.75,$ while confidence degree of both $f_{1}$ to $f_{2}$ is $0.67 .$ That is to say, if the condition is set, the proportion of $f_{2}$ will decrease. This shows that formulae $f_{1}$ and $f_{2}$ is exclusive. Therefore, Promotion Degree is proposed in this paper to tackle this shortcoming.

\textbf{\textit{Definition 5 (Promotion Degree):}} It represents the ratio of confidence degree $p_{Confidence}\left(f_{a} \Rightarrow f_{b}\right)$ to $p\left(f_{b}\right),$ and can be shown as follows:
$$
p_{promotion}\left(f_{a} \Rightarrow f_{b}\right)=\dfrac{p_{Confidence}\left(f_{a} \Rightarrow f_{b}\right)}{p\left(f_{b}\right)}
$$

The promotion degree reflects the correlation between $f_{a}$ and $f_{b}$ in association rule $\left(f_{a} \Rightarrow f_{b}\right)$.
As an empirical statistic over the entire $\mathrm{KG}, p_{promotion}\left(f_{a} \Rightarrow f_{b}\right)$ is larger if more entities with $f_{a}$ also have $f_{b}$ as a neighboring formula. In example above, we regard the ratio of $0.67 / 0.75$ as the promotion degree of $f_{1}$ to $f_{2}$. The higher the promotion degree is, the higher the positive correlation is. The lower the promotion degree is, the higher the negative correlation is. We set a threshold value denoted as $\lambda$. If promotion degree is larger than $\lambda,$ it can be shown that $f_{a}$ and $f_{b}$ are association. If smaller than $\lambda, f_{a}$ and $f_{b}$ is exclusive.

In this paper, we apply the promotion degree of a logic rule to measure the extent of such dependency. As shown in Figure x, the input of this framework is triples of the knowledge graph, and 
the output is the ground rules of different types with corresponding promotion degrees. Moreover, only 
rules with promotion degree greater than a threshold are used in our association rules enhanced knowledge graph embedding.

\begin{figure*}[!t] 
	\setlength{\abovecaptionskip}{0cm}
	\setlength{\belowcaptionskip}{-0.2cm}
	\centering 
	\includegraphics[width=0.75\linewidth]{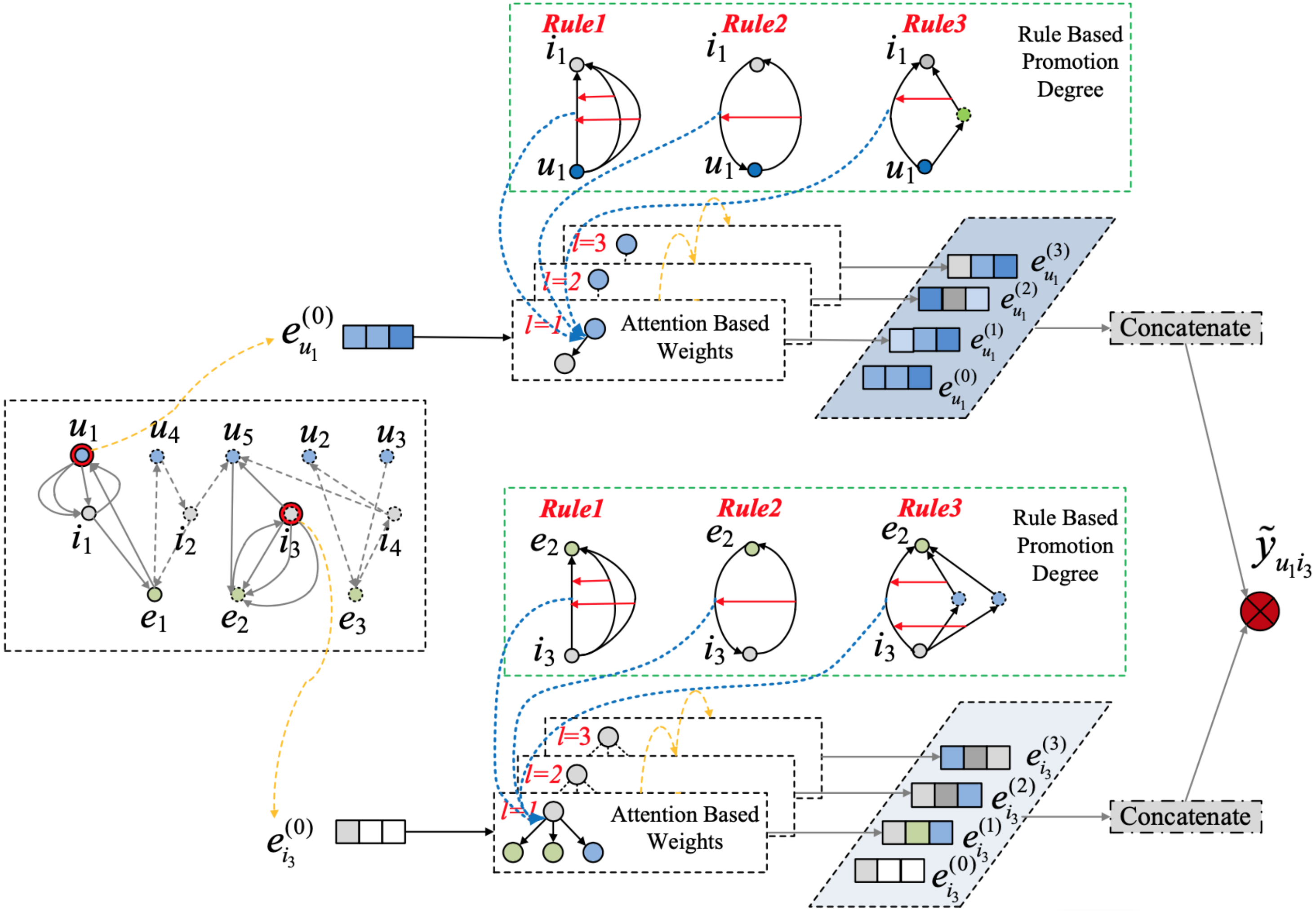} 
	\caption{Illustration of the proposed AR-KGAT model.} 
	\label{arc} 
\end{figure*}

\section{AR-KGAT: Association Rules Enhanced Knowledge Graph Attention Network}

In the attention-based embedding propagation layer, the first-order connectivity information is 
explicitly leveraged to relate representations of entity and relation. Although the performance of GATs 
was proved, they cannot be used for KGs because relation (edge) information and logical rules in the 
neighborhood are neglected. As an integral part of KGs. they could promote more effective aggregation 
of the transformed representations. Unlike the traditional neighborhood aggregation where only 
transformed node representations are used, in the proposed AR-KGAT, node and relation features are 
jointly embedded by exploiting various composition operations from knowledge graph embedding 
techniques, incorporating multi-relational information.

Figure \ref{fig_3} shows an overview of our AR-KGAT framework, where the encoder-decoder structure is adopted. With entity embeddings learned from the encoder module as input, the decoder aims to measure the plausibility of triplets and logical rules. The parameters of the aggressor are adjusted according to the feedback provided by the decoder. The proposed network is trained with the loss function in an end-to-end learning manner. Accordingly, the entity and relation embeddings are learned from ground rules and original triplets.

In the following sub-sections, components of AR-KGAT are described in detail. Specially, we present our injection techniques for the association logical rules into attention aggregator in Section 5.2. Then, the generalization to high-order propagation is described in Section 5.3. In Section 5.4, we introduce our global loss-functions for joint optimization.

\subsection{Model Inputs}
AR-KGAT treats $\mathrm{KB}$ as a graph with multi-typed nodes and relations, denoted as KGs. In a given
KG denoted by $G=(E, R), E$ and $R$ are the entities (nodes) and relations (links), respectively. The $\mathrm{KB}$
contains a set of triplets $K=\left\{\left(e_{i}, r_{k}, e_{j}\right)\right\},$ and each triplet consists of two entities $e_{i}, e_{j} \in E$ and the corresponding relation $r_{j} \in R .$ In the embedding model, representations of entities, relations, and
scoring function $f$ are learned. For a given input triplet $t=\left(e_{s}, r, e_{o}\right), f(t)$ provides the likelihood of $t$ being a valid triplet. Logical rules L are given besides these triplets, including One-to-One and N-to-One association rules introduced in section 4.

The proposed AR-KGAT learns $d$ -dimensional relation representations $h_{r} \in \mathbb{R}^{d}(\forall r \in R)$ along
with entity representations $h_{v} \in \mathbb{R}^{d}(\forall v \in V),$ describing their latent semantics. These two types of
embedding matrics are used as input. The matrix $H \in \mathbb{R}^{N_{e} \times T}$ represents entity embeddings, where the $i^{\mathrm{th}}$ row is the embedding of the entity. $T$ and $N_{e}$ are the dimension of each entity embedding and the
total number of entities, respectively. Similarly, the matrix $G \in \mathbb{R}^{N_{r} \times P}$ represents the relation
embeddings. The outputs of those matrices are $H^{\prime} \in \mathbb{R}^{N_{e} \times T}$ and $G^{\prime} \in \mathbb{R}^{N_{r} \times P}$.

\subsection{Incorporating Neighborhood Attention}
In order to determine the relative weights of neighbors in the attention model, the following two 
problems are considered. (1) The types of neighborhood relations that lead to potentially critical 
neighborhoods. (2) The important neighbors in transformed embedding according to those relations. 
Considering these two requirements, a novel association rules-enhanced attention-based aggregator is 
proposed, incorporating both \textbf{\textit{logic-based}} and \textbf{\textit{graph-based}} attention mechanisms.

Let $N(i)=\left\{\left(e_{i}, r_{k}, e_{j}\right) \mid\left(e_{i}, r_{k}, e_{j}\right) \in G\right\}$ denote the set of neighboring triplets where $e_{i}$ is the head entity. The contribution of the neighboring node $e_{j}$ to $e_{i}^{O}$ is determined according to its relative importance. Traditional methods for aggregating neighborhoods ignored useful information in the
neighbor node $e_{j}$ and high-order relation $r_{k}$ since only the collections of transformed embeddings were considered into account. In order to facilitate more effective aggregation of the transformed embeddings, relation-representations are incorporated into the GAT formulation. The encoder learns the representation of each triplet associated with $e_{i}$ by a linear transformation over the concatenated feature vectors of entity and relation, formulated as follows:
$$
c_{i j k}^{\rightarrow}=W_{1}\left[\vec{h}_{i}\right]\left[\vec{h}_{j}\right]\left[\vec{g}_{k}\right]
$$
where $c_{i j k}^{\rightarrow}$ is the vector form of a triplet $t_{i j}^{k} . \vec{h}_{i}$ and $\vec{h}_{j}$ are the corresponding representations of entities $e_{i}$ and $e_{j},$ respectively. $\vec{g}_{k}$ is the representation of relation $r_{k} \cdot W_{1}$ indicates the linear transformation matrix. This low complexity transformation matrix consists of matrix product operations.

\subsubsection{Association Rule Weighing Mechanism}
With the promotion degree between association rules $\left(f_{a} \Rightarrow f_{b}\right)$ at hand, the neighboring formula
is characterized that it leads to important neighbors. For $\left(f_{a} \Rightarrow f_{b}\right),$ if the logical formula $f_{a}$ has a
large $P_{promotion}\left(f_{a} \Rightarrow f_{b}\right),$ it is statistically relevant to $f_{b} .$ In the proposed model, association relations can be recognized as positive only when a promotion degree is larger than the threshold value $\lambda$.

Under the above intuitions, the logical rule mechanism for measuring the correlation of association rules is implemented as follow:
$$
\alpha_{(i, k, j)}^{L 1}=\prod_{r_{o} \neq r_{k} \& r_{o} \in N_{R}(i)} \log _{\lambda} P_{promotion}\left(rule_{n}(i, j)\right),
$$
where $P_{\text {promotion}}\left(\right.$rule $\left._{n}(i, j)\right) \geq \lambda\ and\ n=\{1,2,3\}$. Specifically,
$$
rule_{1}(i, j)=f_{1}(i, o, j) \Rightarrow f_{1}(i, k, j),
$$
$$
rule_{2}(i, j)=f_{1}(i, o, j) \Leftrightarrow f_{1}(j, k, i),
$$
$$
rule_{3}=f_{2}(i, m, n, j)=r(i, m, k) \wedge r(k, n, j) \Rightarrow f_{1}(i, o, j),
$$
where $f_{1}(i, k, j)$ is a one-hop logical formula with a head entity $e_{i},$ a tail entity $e_{j},$ and body relation
$r_{k} \cdot f_{2}(i, m, n, j)$ is a two-hop logical formula with a head entity $e_{i},$ a tail entity $e_{j},$ and body
relation $r(i, m, k) \wedge r(k, n, j)$.

Note that $\alpha_{(i, k, j)}^{L 1}$ promotes one-hop neighboring formula strongly implying $f_{1}(i, k, j)$ and demotes those implied by some other irrelevant formula in the same neighborhood. Specifically, for any neighboring logical formula $f_{1}(i, o, j)$ that has the larger promotion degree to $f_{1}(i, k, j),$ it is
statistically relevant to $f_{1}(i, k, j)$. The lower the promotion degree is, the higher the negative correlation is. In this paper, only neighboring logical formulae with greater promotion degree than the threshold $\lambda$ are selected in the knowledge graph embedding.

\subsubsection{Neural Network Mechanism}
In our model, which is an extension of classic GAT, different types of relations have different 
weights for aggregation, and the weights are learned during the training. The attention weights are 
guided to be distributed at a coarse granularity of relations with global relations statistics. However, such 
statistics is not enough to consult finer-grained information hidden in the transformed neighbor 
representations to determine the relative importance of neighborhoods. To address this issue, we adopt 
an attention-based mechanism shown in Figure \ref{fig_4}:
\begin{figure}[h]
	\centering 
	\includegraphics[width=\columnwidth]{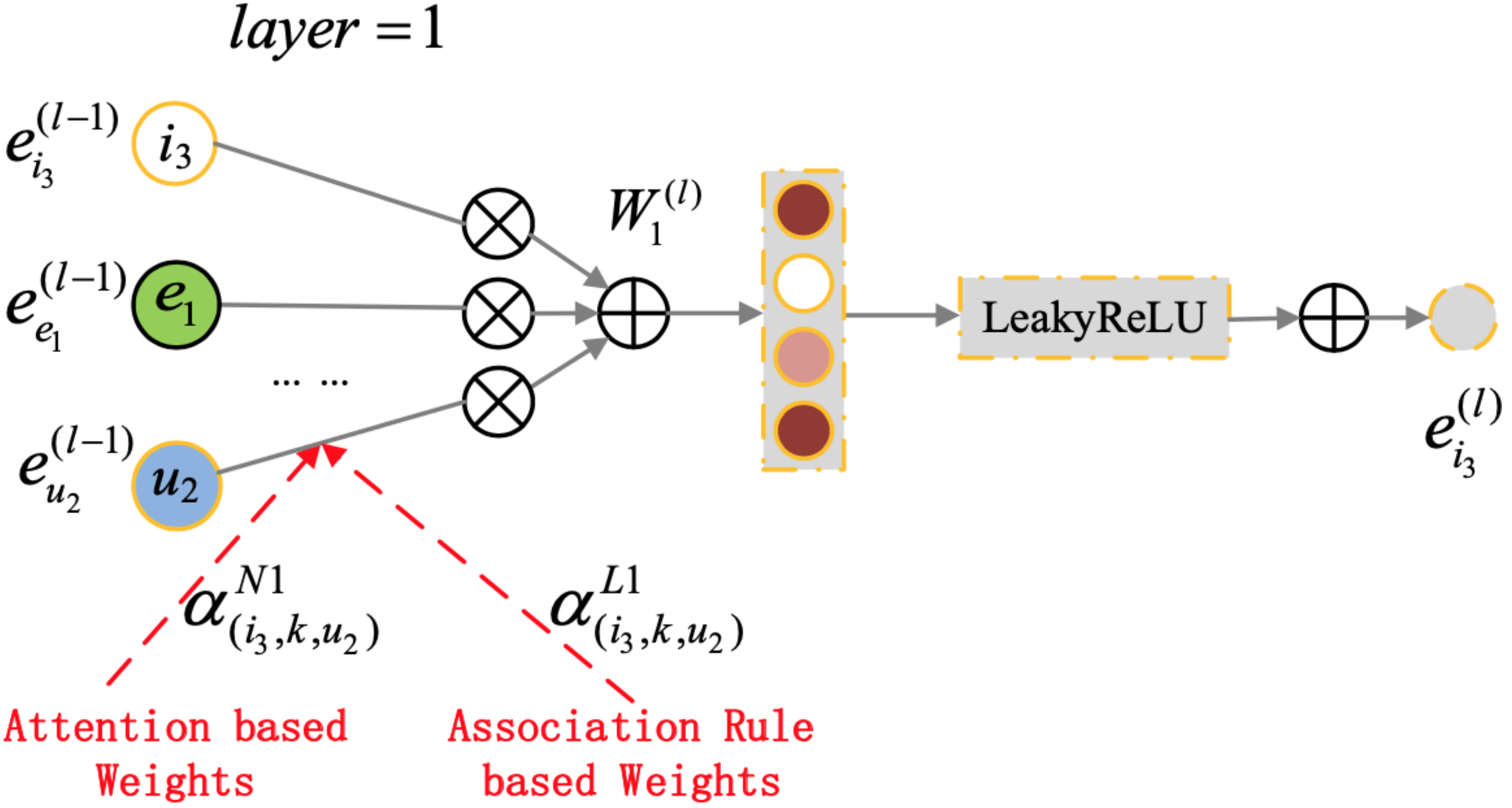}
	\caption{The attentive embedding propagation layer of the proposed model.}
	\label{fig_4}
\end{figure}

Similar to GAT, the proposed attention mechanism is applied to learn the importance of each triplet $b_{(i, j, k)}$ automatically. The linear transformation with a weight matrix $W_{2}$ is conducted, followed by the LeakyRelu activation function. LeakyRelu is employed to obtain the absolute attention value of the 
triplet:
$$
b_{(i, j, k)}=Leaky \operatorname{Re} L U\left(W_{2} \vec{c}_{i j k}\right)
$$

The relative attention values $\alpha_{(i, k, j)}^{N 1}$ for a single triplet are computed via the softmax layer applied 
over $b_{(i, j, k)}$, as shown in the following equation.
$$
\alpha_{(i, k, j)}^{N 1}=soft \max _{j k}\left(b_{(i, k, j)}\right)=\dfrac{\exp \left(b_{(i, k, j)}\right)}{\sum_{j \in N_{i}} \sum_{r_{o} \in R_{i j}} \exp \left(b_{(i, o, j)}\right)}
$$
where $N_{i}$ represents the neighborhood of entity $e_{i} . R_{i j}$ indicates the set of relations of entities $e_{i}$ and $e_{j}$.

Through the softmax function, the coefficients are normalized across all triplets connected with $e_{i}$. Unlike the mechanism of the association rule at the relation-level, the calculation of $\alpha_{(i, k, j)}^{N 1}$ concentrates more on the neighbor feature itself. It also demonstrates that the features of both neighbor entities and relations are helpful in training current triplets. Formally, these two weighting mechanisms are used 
together in the computation of the importance of neighbors, and the new embedding of the entity $e_i$ is the sum of each triplet representation weighted by both logic-based and network-based weights as shown below:
$$
\vec{e}_{i}=\sum_{f_{1}(i, k, j) \in N_{L_{1}}(i)}\left(\alpha_{(i, k, j)}^{L 1}+\alpha_{(i, k, j)}^{N 1}\right) C_{(i, k, j)}^{1}
$$
Here $\alpha_{(i, k, j)}^{L 1}$ and $\alpha_{(i, k, j)}^{N 1}$ are logic-based and network-based attention weights for $e_{j}$ given $e_{i}$ via relation $r_{k}$. This transformation can significantly improve the representation vectors through the 
accumulated and encoded features from local and structured neighborhoods.

\subsection{High-order Propagation}
In the proposed architecture, the notion of the link is extended to a directed formula to consider 
more neighborhood information into the aggregation in sparse graphs. An auxiliary high-order relation is 
introduced for n-hop neighbors between two entities. Representations of all the relations in the logical 
formula are summarized to compute the auxiliary relation. Motivated by the above aggregator 
architecture, an efficient model for high-order propagation is constructed and used to update 
forward-pass any entity. The model is defined as follows:
\begin{figure}[h]
	\centering 
	\includegraphics[width=\columnwidth]{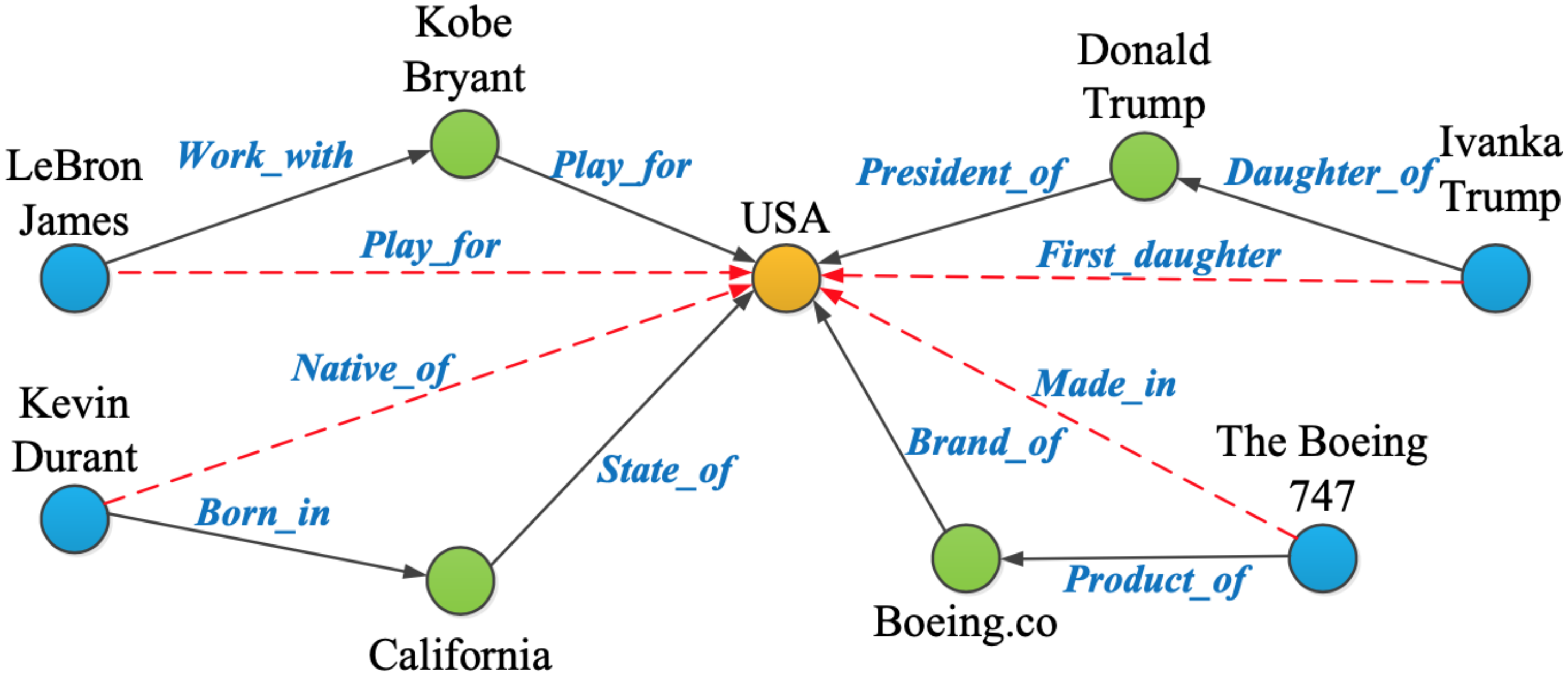}
	\caption{The aggregation process of the graph attention layer. The dashed lines represent an auxiliary link from n-hop neighbors (n=2, here).}
	\label{fig_5}
\end{figure}

Figure \ref{fig_5} shows the first layer of the proposed model, where all entities extract information from their direct in-flowing neighborhoods. The aggregation process learns new representations for entities, and an auxiliary edge among two-hop neighbors is introduced. The USA combines the possessed information about their neighbors in the previous layers from entities Kobe Bryant, California, Boeing.co, and Donald Trump. In the proposed model, knowledge can be iteratively accumulated from distant neighbors guided by a specific logic formula. The propagation layers can be further stacked to obtain more higher-order connectivity, collecting the information about higher-hop neighborhoods. Note that the influence of distant entities is exponentially decreased according to the increased model depth. 

Furthermore, the initially embedded information of entities gets lose over the learning new embeddings. To resolve this problem, $H^{t}$ is obtained by a linear transformation with a weight matrix
$W^{E} \in \mathbb{R}^{T^{i} \times T^{f}}$ from the input entity embeddings $H^{i} . H^{t}$ represents the transformed entity embeddings. $T^{i}$ and $T^{f}$ are the dimensions of the initial and final entity embeddings, respectively. 
The initial entity embedding is added to the entity embeddings obtained in the final attention layer, as follows:
$$
H^{\prime \prime}=W^{E} H^{t}+H^{f}
$$

Considering the computational complexity, we use one-hop to three-hop neighborhood information 
for the propagation in this paper. The final attention values indicate the relative significance of 
neighboring nodes to obtain collaborative signals. The attention parameters in the encoder are learned in 
the training process using label data. Unlike the information propagation in GCN and GAT, our method 
exploits the association rules of KB and specifies the different significance of neighbors.

\subsection{Global Objective Function}
The decoder measures the plausibility of the trained triplet and logical rules, given the entity and 
relation representations output from the encoder. Triplets and logical rules are unified as an atomic and 
complex formula. Then, representations for entity and relation are learned by minimizing the global loss, 
optimizing the objective function.

\subsubsection{Truth Function}
A training set $F$ contains all positive samples, including: (i) observed triplets; (ii) ground rules. Further, a truth function $I: F \rightarrow[0,1]$ assigns a soft truth value to each formula, indicating the degree that ground-rule is met or triplet holds.

Also, t-norm fuzzy logic is used to model the rules. It decomposes a complex formula into its constituents via specific t-norm based logical connectives. Moreover, we define the compositions
associated with logical negation $(\neg),$ conjunction $(\wedge),$ and disjunction $(\vee)$:
$$
I\left(f_{1} \wedge f_{2}\right)=I\left(f_{1}\right) \cdot I\left(f_{2}\right)
$$
$$
I\left(f_{1} \vee f_{2}\right)=I\left(f_{1}\right)+I\left(f_{2}\right)-I\left(f_{1}\right) \cdot I\left(f_{2}\right)
$$
$$
I\left(\neg f_{1}\right)=1-I\left(f_{1}\right)
$$
where $f_{1}$ and $f_{2}$ are two constituent formula.

Then, the truth values of the constituent triples can be used to determine the truth values of two 
types of association rules using the corresponding logical connectives:

(1) Given a ground one-to-one association rule:

For the inference rule $L_{1 \rightarrow 1(\text { inf } e r)}\triangleq\left(e_{m}, r_{s}, e_{n}\right) \Rightarrow\left(e_{m}, r_{t}, e_{n}\right),$ the truth value is calculated as:
$$
I\left(L_{1 \rightarrow 1(\text { inf } e r)}\right)=I\left(e_{m}, r_{s}, e_{n}\right) \cdot I\left(e_{m}, r_{t}, e_{n}\right)-I\left(e_{m}, r_{s}, e_{n}\right)+1
$$

Also, for the anti-symmetry rule $L_{1 \rightarrow 1(a n t i)}\triangleq\left(e_{m}, r_{a}, e_{n}\right) \Leftrightarrow\left(e_{n}, r_{b}, e_{m}\right)$, the truth value is 
calculated as:
$$
I\left(L_{1 \rightarrow 1(anti)}\right)=I\left(e_{m}, r_{a}, e_{n}\right) \cdot I\left(e_{n}, r_{b}, e_{m}\right)-I\left(e_{m}, r_{a}, e_{n}\right)+1
$$
where the truth value of a constituent triple $I(\cdot, \cdot, \cdot)$ is defined as:
$$
I\left(e_{i}, r_{k}, e_{j}\right)=1-\dfrac{1}{3 \sqrt{d}}\left\|\vec{e}_{i}+\vec{r}_{k}-\vec{e}_{j}\right\|_{1}
$$
$\vec{e}_{i}, \vec{e}_{j}$ and $\vec{r}_{k}$ are vector representations of corresponding entities $e_{i}, e_{j},$ and relation $r_{k}$.

(2) Given a ground two-to-one association rule:
$$
L_{2 \rightarrow 1}\triangleq\left(e_{l}, r_{s_{1}}, e_{m}\right) \wedge\left(e_{m}, r_{s_{2}}, e_{n}\right) \Rightarrow\left(e_{l}, r_{t}, e_{n}\right),
$$
The truth value is calculated by:
$$
\begin{aligned}
I\left(L_{2 \rightarrow 1}\right)=&I\left(e_{l}, r_{s_{1}}, e_{m}\right) \cdot I\left(e_{m}, r_{s_{2}}, e_{n}\right) \cdot I\left(e_{l}, r_{t}, e_{n}\right)\\ &-I\left(e_{l}, r_{s_{1}}, e_{m}\right) \cdot I\left(e_{m}, r_{s_{2}}, e_{n}\right)+1
\end{aligned}
$$

A larger truth value indicates that ground rules are more-satisfied. In our proposed AR-KGAT 
framework, any rules represented as the first-order logic formula can be handled besides these two types 
of rules.

\subsubsection{Loss Function}
The training dataset consists of the original triplets and two types of ground rules. The dataset is 
used to train the proposed model, where the global loss is minimized to learn the entity and relation 
embedding representation. The positive formula is forced to have larger truth values than negative ones:
$$
\min _{\{\mathrm{e}\},\{\mathrm{r}\}} \sum_{f^{+} \in F} \sum_{f^{-} \in N_{f+}}\left[\gamma-I\left(f^{+}\right)+I\left(f^{-}\right)\right]_{+}
$$
$f^{+} \in F$ represents positive samples, including triplets and ground rules. $f^{-} \in N_{f+}$ represents
negative samples constructed by corrupting $f^{+} . \gamma>0$ is a separate-parameter of positive and
negative samples. The constraints that $\forall e \in E,\|e\|_{2} \leq 1, \forall r \in R,\|r\|_{2} \leq 1$ are employed during the training process.

(1) For positive triplet $f^{+}, \gamma>0$ being a triplet sample $\left(e_{m}, r_{k}, e_{n}\right),$ the $f^{-}$ is generated by
replacing one of the head and tail entities by a random entity. The procedure is defined as follows:
$$
N_{f^{+}}=\left\{\left(e_{m}^{\prime}, r_{k}, e_{n} \mid e_{m}^{\prime} \in E\right) \cup\left(e_{m}, r_{k}, e_{n}^{\prime} \mid e_{n}^{\prime} \in E\right)\right\}
$$
The triplets from the knowledge graph $\Delta$ are considered as positive triplets. For each $(s, q, o) \in \Delta$,
the negative triplets $\Delta_{(s, q, o)}^{\prime}$ are generated by randomly corrupting one of the object and subject entities
by another entity in $E .$ For instance, a negative instance (Berlin, Capital-Of, France) can be generated from the positive triplet (Berlin, Capital-Of, Germany).

(2) For positive rule $f^{+}:$ ground-rule sample of one-to-one association rule

For inference rule $L_{1 \rightarrow 1(\text { inf } e r)}\triangleq\left(e_{m}, r_{s}, e_{n}\right) \Rightarrow\left(e_{m}, r_{t}, e_{n}\right),$ one of the head and tail entities is replaced by a random one to generate the negative rule $f^{-},$ as described in the following:
$$
\begin{aligned}
N_{f^{+}}=\left\{\left[\left(e_{m}^{\prime}, r_{s}, e_{n}\right)\right.\right. &\left.\Rightarrow\left(e_{m}^{\prime}, r_{t}, e_{n}\right) \mid e_{m}^{\prime} \in E\right] \cup\left[\left(e_{m}, r_{s}, e_{n}^{\prime}\right)\right.\\ 
&\left.\left.\Rightarrow\left(e_{m}, r_{t}, e_{n}^{\prime}\right) \mid e_{n}^{\prime} \in E\right]\right\}
\end{aligned}
$$

For an anti-symmetry rule $L_{1 \rightarrow 1(a n t i)}\triangleq\left(e_{m}, r_{a}, e_{n}\right)\Leftrightarrow\left(e_{n},\right.$ $\left.r_{b}, e_{m}\right)$, one of the head and tail entities is 
replaced by a random one to generate the negative rule $f^{-}$, as described in the following:
$$
\begin{aligned}
N_{f^{+}}=\left\{\left[\left(e_{m}^{\prime}, r_{a}, e_{n}\right)\right.\right. &\left.\Rightarrow\left(e_{m}^{\prime}, r_{b}, e_{n}\right) \mid e_{m}^{\prime} \in E\right] \cup\left[\left(e_{m}, r_{b}, e_{n}^{\prime}\right)\right.\\ 
&\left.\left.\Rightarrow\left(e_{m}, r_{a}, e_{n}^{\prime}\right) \mid e_{n}^{\prime} \in E\right]\right\}
\end{aligned}
$$

(3) For positive rule $f^{+}:$ ground-rule sample of n-to-one association rule
$$
L_{2 \rightarrow 1}\triangleq\left(e_{l}, r_{1}, e_{m}\right)+\left(e_{m}, r_{2}, e_{n}\right) \Rightarrow\left(e_{l}, r_{3}, e_{n}\right)
$$
One of the $e_{l}$ and $e_{n}$ is replaced by a random entity to generate the negative rule $f^{-}$ according, as described in the following:
$$
\begin{aligned}
N_{f^{+}}=&\left\{\left[\left(e_{l}^{\prime}, r_{1}, e_{m}\right)+\left(e_{m}, r_{2}, e_{n}\right) \Rightarrow\left(e_{l}^{\prime}, r_{3}, e_{n}\right) \mid e_{l}^{\prime} \in E\right] \cup\right.\\ 
&\left.\left[\left(e_{l}, r_{1}, e_{m}\right)+\left(e_{m}, r_{2}, e_{n}^{\prime}\right) \Rightarrow\left(e_{l}, r_{3}, e_{n}^{\prime}\right) \mid e_{n}^{\prime} \in E\right]\right\}
\end{aligned}
$$

The generated negative samples, including corrupted triplets and rules, $f^{-}$ are different from the original triplets and ground rules. The generated negative rules are used as negative samples during the training process, leading to lower values for the negative rules. Note that the embedding representation 
remains the same for an entity that appears in the head position and tail position of a triple in this paper.

\subsubsection{Optimization}
The training of the proposed model is conducted as a two-step procedure. First, the specially 
designed GAT is trained to encode the graph entities and relations information. Then, a decoder model is 
trained to perform specific tasks, such as relation prediction task. The Stochastic Gradient Descent (SGD) 
algorithm is used with a mini-batch to optimize the minimization problem. In the proposed AR-KGAT 
framework, the global loss is minimized over the training formula $F$. Accordingly, the embedding 
representations for entity and relation are learned compatible with both triplets and rules. The objective 
function provides lower values for positive samples than for negative samples. The set of positive 
triplets and formula are randomly traversed multiple times. The gradient is back-propagated to update 
the corresponding model parameters after each mini-batch process.

\section{Experimental Evaluation}
The performance of the proposed model AR-KGAT, especially the embedding propagation layer, 
is evaluated on three real-world datasets in typical tasks of knowledge graph completion: the link 
prediction and triplet classification. In this section, first, the performance of the proposed AR-KGAT is 
compared with state-of-the-art knowledge graph embedding methods. Second, the contribution analysis 
of each component of the proposed AR-KGAT, including association rules, attention mechanism, and 
aggregator selection, are also analyzed. Lastly, the effects of the hyper-parameters of the proposed 
AR-KGAT are analyzed.

\subsection{Benchmark Datasets}
In the experiments, the performance of the proposed model is evaluated on two widely used public datasets, WN18RR \cite{34} and FB15k-237 \cite{35}, for link prediction and triplet classification tasks.

\begin{table*}
	\footnotesize
	\setlength{\abovecaptionskip}{0cm}
	\setlength{\belowcaptionskip}{0.2cm}
	\centering
	\caption{Statistics of the datasets.}
	\label{data}
	\setlength{\tabcolsep}{1mm}{
	\begin{tabular}{|c|c|c|c|c|c|c|}
		\hline
		\textbf{Dataset}   & \textbf{Entities} & \textbf{Relations} & \textbf{Training} & \textbf{Validation} & \textbf{Test} & \textbf{Total} \\ \hline
		\textbf{FB15k-237} & 14541             & 237                & 272115            & 17535               & 20466         & 310116         \\ \hline
		\textbf{WN18RR}    & 40943             & 11                 & 86835             & 3034                & 3134          & 93003          \\ \hline
	\end{tabular}
	}
\end{table*}

\textbf{FB15k-237.} FB15K \cite{36} includes the subset of knowledge base triplets, originally derived from 
Freebase. The dataset consists of 310,116 triplets with 14,541 entities, 237 relations. As shown in Table 
1, the dataset is randomly split. The FB15k-237 is consists of textual mentions of Freebase entity pairs 
and knowledge base relation triplets \cite{35}.

\textbf{WN18RR.} WN18RR is generated from WN18RR \cite{36}. As a subset of WordNet, WN18RR
includes 40,943 entities and 18 relations. The text triplets generated by inverting triplets in WN18RR are 
removed to construct the WN18RR dataset to ensure no inverse relation test leakage exists in the 
evaluation dataset. In summary, 93,003 triplets composed of 40,943 entities and 11 relation types are 
used in the WN18RR dataset.

Table 1 summarizes the detailed information on WN18RR and FB15k-237. The triplets samples of 
each dataset are divided into training, validation, and test sub-datasets. They are utilized in the model 
training, hyper-parameter tuning for model selection, and evaluating the model. WN18RR is split following the original data split, while only triplets associated with the 237 relations are employed from each training, validation, and test datasets of FB15K.
\begin{table*}
	\centering
	\caption{Examples of rules created}
	\label{tab_4}
	\begin{tabular}{|c|c|}
		\hline
		\textbf{Dataset}                    & \textbf{Rule examples}                                                                                                                                                                                                  \\ \hline
		\multirow{6}{*}{\textbf{FB15K-237}} & $\forall x, y:$ / sports / athlete / team $(x, y) \Rightarrow$ /sports / sports \_team / player $(y, x)$                                                                                                                \\ \cline{2-2} 
		& \begin{tabular}[c]{@{}c@{}}$\forall x, y, z: /$ people $/$ person $/$ nationality $(x, y) \wedge$ /location / country / offical \_language $(y, z)$\\ $\Rightarrow$ / people / person / language $(x, z)$\end{tabular}  \\ \cline{2-2} 
		& \begin{tabular}[c]{@{}c@{}}$\forall x, y, z:$ country $/$ ad min istrative divisions $(x, y) \wedge$ /ad min istrative division $/$ capital $(y, z)$\\ $\Rightarrow$ / people / person / language $(x, z)$\end{tabular} \\ \cline{2-2} 
		& $\forall x, y:$ llocation $/$ country $/$ capital $(x, y) \Rightarrow$ location / location $/$ contains $(x, y)$                                                                                                        \\ \cline{2-2} 
		& $\forall x, y:$ /location / country / first \_level $_{-}$ divisions $(x, y) \Rightarrow$ location / location $/$ contains $(x, y)$                                                                                     \\ \cline{2-2} 
		& $\forall x, y:$ people / family / members $(x, y) \Rightarrow$ / people / person / place $_{-}$ of $_{-}$ birth $(y, x)$                                                                                                \\ \hline
		\multirow{4}{*}{\textbf{WN18RR}}    & $\forall x, y:_{-}$ member $_{-}$ of $_{-}$ domain $_{-}$usage$(x, y) \Leftrightarrow_{-}$ synset \_domain\_usage\_of $(y, x)$                                                                                          \\ \cline{2-2} 
		& $\forall x, y:_{-}$ hyponym $(x, y) \Leftrightarrow_{-}$ hupernym $(y, x)$                                                                                                                                              \\ \cline{2-2} 
		& $\forall x, y:_{-}$ part $_{-}$ of $(x, y) \Leftrightarrow_{-}$ has $_{-} \operatorname{part}(y, x)$                                                                                                                    \\ \cline{2-2} 
		& $\forall x, y:_{-}$ins tan ce\_hyponym$(x, y) \Leftrightarrow_{-}$ins tan ce\_hupernym$(y, x)$                                                                                                                          \\ \hline
	\end{tabular}
\end{table*}

First, the two types of association rules are constructed for each dataset, in the form of Rule 1 (one-to-one inference rule $\forall x, y:\left(x, r_{s}, y\right) \Rightarrow\left(x, r_{t}, y\right)$), Rule 2 (one-to-one anti-symmetric rule $\forall x, y:\left(x, r_{s}, y\right) \Leftrightarrow\left(y, r_{t}, x\right)$),  and Rule 3 ( n-to-one transitivity rule $\forall x, y, z:\left(x, r_{s_{1}}, y\right) \wedge\left(x, r_{s_{2}}, z\right) \Rightarrow\left(x, r_{t}, z\right)$) following the schemes described in Section 4. The entity and relation embeddings are obtained by using the proposed aggregator. Then, the truth values for each rule are computed.
\begin{table}
	\centering
	\caption{The rule statistics.}
	\label{tab_2}
	\begin{tabular}{|c|c|c|c|c|}
		\hline
		\textbf{Dataset} & \textbf{\#Rule 1} & \textbf{\#Rule 2} & \textbf{\#Rule 3} & \textbf{Total} \\ \hline
		FB15K-237        & 160               & 2357              & 637               & 3334           \\ \hline
		WN18RR           & 0                 & 11                & 0                 & 11             \\ \hline
	\end{tabular}
\end{table}

Table \ref{tab_2} summarizes the details of the mined rules. For the FB15K-166 and FB15K-237 datasets,
the threshold $\tau$ is set 0.5 for rule $1,$ rule 2 and rule $3 .$ For the WN18RR dataset, the threshold $\tau$ is also set as 0.5 for rule 2. The threshold values were determined using the validation datasets. The method illustrated in Section 3 instantiates the mined rules with concrete entities (grounding) after the association rules mining.
\begin{table}
	\centering
	\caption{The candidates of ground rule.}
	\label{tab_3}
	\begin{tabular}{|c|c|c|c|c|}
		\hline
		\textbf{Dataset} & \textbf{\#Rule 1} & \textbf{\#Rule 2} & \textbf{\#Rule 3} & \textbf{Total} \\ \hline
		FB15K-237        & 8508               & 152987              & 113476               & 274971           \\ \hline
		WN18RR           & 0                 & 9352                & 0                 & 9352             \\ \hline
	\end{tabular}
\end{table}

Table \ref{tab_3} and Table \ref{tab_4} present the candidates of the ground rules and some examples of the rules in 
both datasets, respectively.

\subsection{Training Details}
Two sets of negative samples of triplets are generated, where the head entity is randomly replaced 
in the first set, while the tail entity is randomly replaced in the second set. Equal numbers of negative 
triplets from two sets are generated to achieve more robust performance in detecting both head and tail. 
The embeddings for entity and relation are initialized with TransE. The default Xavier initializer is 
applied to initialize the model parameters. The hyper-parameter ranges are specified as follows: learning 
rate \{0.01,0.005,0.003,0.001\}, dropout-rate \{0, 0.1, 0.2, 0.3, 0.4, 0.5\}, embedding dimension {50, 100, 
	150, 200, 250}, and number of neighbors \{3, 4, 5, 6, 7, 8, 9, 10, 11\}. We use the adaptive moment 
(Adam) algorithm to optimize the model weights. A grid search is used to determine the 
hyper-parameters in our models during the training. Also, an early training-termination strategy is 
employed, \textit{i.e.}, the training process is prematurely terminated when Hit@k on the validation set is not 
increased during 100 successive epochs. Different hyper-parameters are evaluated on the validation 
dataset to select the optimal model. The optimal hyper-parameters are described in Section 6.7. For each 
dataset, the following settings are used in this experiment. For the FB15k-237, the dropout rate is 0.2, 
the maximum number of neighbor relations is 8, the learning rate is 0.003, and the embedding size is 
100; for the WN18RR dataset, the dropout rate is 0.2, the maximum number of neighbor relations is 6, 
the learning rate is 0.003, and the embedding size is 150. Moreover, each dataset is split into three sets 
for training, validation, and testing, which is the same as the setting of the original models. The proposed 
model was implemented in PyTorch, and the experiments were conducted on two NVIDIA GTX 2080Ti GPU. The computational times for one epoch are 141 seconds and 15 seconds for the FB15k-237 and WN18RR datasets, respectively.

\subsection{Baselines}
The proposed model is evaluated with comparisons to the state-of-the-art methods categorized as follows:

$\bullet$ \textbf{Translation-based methods:} Relatively simple vector-based methods. For example, TransE\cite{10}, 
TransR\cite{11}, DistMult\cite{13} and ComplEx\cite{14}.

$\bullet$ \textbf{Deep Neural network-based methods:} A series of deep non-linear neural network-based 
methods, including convolution based models (including ConvE\cite{18}, ConvKB\cite{33} and ConvR\cite{40}), and 
graph neural network based models (including R-GCN\cite{20}, CompGCN\cite{38} and SACN\cite{39}).

$\bullet$ \textbf{Logical rules powered methods:} Several methods which jointly embed knowledge graphs and 
logical rules, including LR-KGE\cite{32} and KALE\cite{37}.

$\bullet$ \textbf{Variant of our model:} AR-KGCN is referred to as a variant of our proposed AR-KGAT, which 
contains GCN based aggregator without consideration of attention mechanism.

Among these compared methods, except for LR-KGE, KALE, AR-KGCN, and AR-KGAT, other baselines use only triplets. In the experiments, the effectiveness of the proposed model is evaluated with state-of-the-art methods on two typical tasks for knowledge graph completion: link prediction (Section 6.4) and triplet classification (Section 6.5).
\begin{table*}
	\centering
	\caption{ Results of link prediction on the testing dataset of FB15k-237 and WN18RR.}
	\label{tab_5}
	\begin{tabular}{|c|c|c|c|c|c|c|c|c|}
		\hline
		\multirow{2}{*}{\textbf{Methods}} & \multicolumn{4}{c|}{\textbf{FB15K-237}}                                                                                            & \multicolumn{4}{c|}{\textbf{WN18RR}}                                                                                               \\ \cline{2-9} 
		& \textbf{MRR} & \textbf{$\underline{\text{Hits}@1}$} & \textbf{$\underline{\text{Hits}@3}$} & \textbf{$\underline{\text{Hits}@10}$} & \textbf{MRR} & \textbf{$\underline{\text{Hits}@1}$} & \textbf{$\underline{\text{Hits}@3}$} & \textbf{$\underline{\text{Hits}@10}$} \\ \hline
		\textbf{TransE}                   & 0.224        & 0.138                                & 0.246                                & 0.395                                 & 0.236        & 0.427                                & 0.441                                & 0.532                                 \\ \hline
		\textbf{ComplEX}                  & 0.247        & 0.158                                & 0.287                                & 0.428                                 & 0.392        & 0.410                                & 0.463                                & 0.510                                 \\ \hline
		\textbf{DisMult}                  & 0.240        & 0.163                                & 0.267                                & 0.422                                 & 0.403        & 0.391                                & 0.440                                & 0.496                                 \\ \hline
		\textbf{ConvE}                    & 0.317        & 0.237                                & 0.352                                & 0.493                                 & 0.458        & 0.400                                & 0.440                                & 0.525                                 \\ \hline
		\textbf{ConvKB}                   & 0.396        & 0.198                                & 0.324                                & 0.517                                 & 0.465        & 0.058                                & 0.441                                & 0.558                                 \\ \hline
		\textbf{ConvR}                    & 0.335        & 0.242                                & 0.401                                & 0.511                                 & 0.466        & 0.432                                & 0.494                                & 0.529                                 \\ \hline
		\textbf{SACN}                     & 0.354        & 0.269                                & 0.393                                & 0.547                                 & 0.470        & 0.435                                & 0.486                                & 0.548                                 \\ \hline
		\textbf{CompGCN}                  & 0.355        & 0.264                                & 0.397                                & 0.535                                 & 0.479        & 0.443                                & 0.494                                & 0.546                                 \\ \hline
		\textbf{R-GCN}                    & 0.419        & 0.311                                & 0.434                                & 0.558                                 & 0.483        & 0.438                                & 0.497                                & 0.567                                 \\ \hline
		\textbf{KALE}                     & 0.282        & 0.271                                & 0.289                                & 0.464                                 & 0.376        & 0.222                                & 0.348                                & 0.489                                 \\ \hline
		\textbf{LR-KGE}                   & 0.338        & 0.241                                & 0.346                                & 0.533                                 & 0.463        & 0.430                                & 0.522                                & 0.528                                 \\ \hline
		\textbf{AR-KGCN}                  & 0.360        & 0.287                                & 0.404                                & 0.555                                 & 0.476        & 0.428                                & 0.516                                & 0.571                                 \\ \hline
		\textbf{AR-KGAT}                  & 0.442        & 0.361                                & 0.483                                & 0.581                                 & 0.518        & 0.465                                & 0.540                                & 0.626                                 \\ \hline
	\end{tabular}
\end{table*}

\subsection{Link Prediction}
The task of link prediction is formulated to predict missing facts using the given facts in Knowledge Graphs. A directed labeled graph $G=(V, E, R)$ is used to represent the knowledge base. However, in the link prediction task, an incomplete subset $\hat{E}$ is given rather than the full set of edges $E .$ Thus, the link prediction is stated as completing a triplet $\left(e_{i}, r_{k}, e_{j}\right)$ when $e_{i}$ or $e_{j}$ missing, i.e.
the prediction of $e_{j}$ given $\left(e_{i}, r_{k}, ?\right)$ or the prediction of $e_{i}$ given $\left(?, r_{k}, e_{j}\right) .$ The task is to assign scores $f\left(e_{i}, r_{k}, e_{j}\right)$ to possible edges $\left(e_{i}, r_{k}, e_{j}\right)$ to determine the degree to which those edges belong to $E$.

\textbf{Implementation:} Each entity $e_{i}$ is replaced by every other entity $e_{i}^{\prime} \in E \backslash e_{i}$ to construct a set of $(N-1)$ corrupt triplets for each test triple $\left(e_{i}, r_{k}, e_{j}\right) .$ Then, a score is assigned to each corrupted triple $\left(e_{i}^{\prime}, r_{k}, e_{j}\right) .$ The correct entity $e_{i}$ is ranked in descending order of scores (or in ascending order of the distances). The rank for the corrupted entity of the tail entity $e_{j}$ is obtained in a similar way. All the models are evaluated in a filtered setting, where the corrupted triplets already presented in other datasets 
are removed during the ranking process.

\textbf{Evaluation protocol:} We employ the widely used two metrics for evaluation: mean reciprocal rank (MRR) and Hits@k (the proportion of ground truth entities ranked top- $k \in\{1,3,10\}$ ). Higher values indicate better performance for both MRR and Hits@k. Also, the filtered setting (Bordes et al., 2013) is used because the knowledge graphs contain some corrupted triplets. In other words, all valid triplets are filtered out before ranking.

The results of link prediction on FB15K and WN18RR datasets are shown in Table 5. The results clearly indicate that our proposed method outperforms state-of-the-art methods in terms of four metrics, especially for the Hit@k metrics. Our logical rule-enhanced model greatly improves the performance over the triplets-only-based methods. On average, our AR-KGAT obtains an improvement of at least 9\% and 7.5\% on FB15k-237 and WN18RR in terms of MRR over other baselines. Also, the Hits@1 is increased by 0.282 with our AR-KGAT model on the WN18RR dataset, showing the promising potential of the proposed method. Note that, among different metrics, the first rank metric is the most essential in 
practical knowledge inference applications. Low values of Hits@1 usually refer that the corrected triplet is not likely to be ranked in the first place. From this point of view, the proposed AR-KGAT model can provide knowledge graph embedding for practical applications of knowledge inference.

Furthermore, the proposed AR-KGAT model improves the performance for the FB15k-237 test dataset, Hits@10 by a margin of 4.1\%, and Hits@3 by a margin of 5.7\%. Also, the performance of the 
proposed model is improved by 8.3\% and 9.3\%, respectively, in terms of Hits@10 and Hits@3 on the 
WN18RR test dataset. The logical rule-enhanced models, including LR-KGE and AR-KGAT, mostly 
outperform the original models, where triplets are used alone on both datasets. This implies that 
integrating logical rules can indeed infer new and corrected triplets. Specifically, compared to 
GNN-based methods, our AR-KGAT model outperforms on both benchmark datasets in terms of all the 
evaluation metrics. It validates the superiority of our model to other aggregators and the necessities to 
integrate the association rule-based weighting mechanism into the overall framework. Moreover,
AR-KGCN consistently underperforms AR-KGAT, which illustrates that removing attention components 
degrades the performance of the model. On the whole, our rule-enhanced GAT framework outperforms 
other compared methods consistently. It shows that the entities and relations are efficiently represented 
in a low-dimensional space, leveraging the logical rules and attention mechanism.

\subsection{Triple Classification}
The discriminative ability to distinguish between true and false facts is evaluated in a classical 
triplet classification task in knowledge graph embedding. Triplet classification is formulated to determine if a given triplet $(h, r, t)$ is correct or wrong. The validation and testing datasets include both positive and negative triplets in the benchmark datasets, while the training set contains only positive
triplets. We use the thresholds $\delta_{r}$ for each relation to address this task. It is optimized by maximizing the classification accuracy in all triplets with the corresponding relation on the validation set. A triplet $(h,r, t)$ is classified as positive if the dissimilarity score $\phi(h, r, t) \geq \delta_{r}$; otherwise, it is classified as negative. 

\textbf{Implementation:} The same datasets used in the link prediction task are used in triplet classification. 
Accordingly, the ranges of hyper-parameters are also the same as the link prediction. Thus, the optimal 
hyper-parameters are also used without any change for the triplet classification. The estimated truth 
values (or distances) are used to classify the triplets. In this way, a triplet having a large truth value 
(equivalently small distance) is likely to be determined as positive. The triplets association with each 
specific relation are ranked in descending order of the truth values (in ascending order of the distances) 
in the evaluation. The precisions are averaged for each relation. Then, the mean average precision (MAP)
is computed over different relations in test sets.
\begin{table}
	\centering
	\caption{The accuracy comparison of triplet classification for the FB15k-237 and WN18RR test sets.}
	\label{tab_6}
	\begin{tabular}{|c|c|c|}
		\hline
		\textbf{Dataset} & \textbf{FB15K-237} & \textbf{WN18RR} \\ \hline
		\textbf{TransE}  & 0.8258             & 0.9470          \\ \hline
		\textbf{TransR}  & 0.8196             & 0.9513          \\ \hline
		\textbf{ConvE}   & 0.8986             & 0.9686          \\ \hline
		\textbf{ConvKB}  & 0.8890             & 0.9601          \\ \hline
		\textbf{ConvR}   & 0.9094             & 0.9780          \\ \hline
		\textbf{CompGCN} & 0.8991             & 0.9523          \\ \hline
		\textbf{R-GCN}   & 0.9068             & 0.9726          \\ \hline
		\textbf{SACN}    & 0.9011             & 0.9679          \\ \hline
		\textbf{KALE}    & 0.8856             & 0.9573          \\ \hline
		\textbf{LR-KGE}  & 0.8993             & 0.9646          \\ \hline
		\textbf{AR-KGCN} & 0.9104             & 0.9527          \\ \hline
		\textbf{AR-KGAT} & 0.9254             & 0.9733          \\ \hline
	\end{tabular}
\end{table}

\textbf{Results:} Table \ref{tab_6} compares the accuracy of the triplet classification for the FB15k-237 and WN18RR datasets. The analytical results of the triplet classification show consistent observation as the results obtained in the link prediction task. The CNN and GNN based models, including CompGCN, R-GCN, and ConvKB, slightly outperform the original TransE and TransKB models on both datasets in most cases. For these results, we conclude that the translational characteristic between relation and entity can be effectively kept in CNN or GNN based models using the neural network, achieving promising performance in the triplet classification. Also, adopting the logical rule into the models shows significant improvements over the naïve methods that utilize triplets only. This shows that the knowledge representation, such as entity and relation embedding, can be learned better in the translation-based knowledge graph embedding models by the proposed rule-enhanced approach. Further, more rule-related information is exploited in the proposed rule-powered model compared to adding an inferred triplet into training data. The best average accuracy is achieved by the proposed AR-KGAT, confirming the advantages of the proposed model over compared methods in the triplet classification. This observation jointly validates with the observation in Section 6.4 that more predictive embeddings and extended ability beyond the capability of pure logical inference can be learned in a joint embedding scenario. The importance of the attention mechanism is proved by the consistently superior performance of the AR-KGAT over the AR-KGCN, showing it is one of the key components of the proposed model. Equal weighting for all neighbors regardless of their importance might lead to worse embedding propagation process.  

In the following subsections, the proposed AR-KGAT is further analyzed in terms of generalization ability to other configurations. Also, whether it outperforms the others for expected reasons is analyzed.

\subsection{Study of AR-KGAT Model}
To get deep insights into the AR-KGAT, we study the influences of logical rules-based mechanisms and different types of logical rules on the performance. Then, the convergence speed of our proposed AR-KGAT is analyzed. 

\subsubsection{The analysis of the association rule-based mechanism}
In this experiment, the contribution of the logical rules is analyzed on the improvement of the graph attention network in our model. To this end, we conduct further experiments that a single model of the only logical rule mechanism (so-called Logical rules Only) is evaluated. Specifically, the proposed AR-KGAT is compared with three degenerated methods, including AR-KGAT-OptOnly, AR-KGAT-AggOnly, and AR-KGAT-TriOnly. In AR-KGAT-OptOnly and AR-KGAT-AggOnly, the logical rule mechanism is removed from the aggregator and objective function, respectively. As a basic graph attention network framework, AR-KGAT-TriOnly uses only triplets in the optimization function and aggregator to conduct the embedding task. Moreover, AR-KGAT further incorporates both training triplets and ground rules before and during embedding in a joint learning framework. 
\begin{table*}
	\centering
	\caption{Link prediction results by relation category for LR-KGE and AR-KGAT on FB15k-237. The complex relations are more effectively captured in the proposed AR-KGAT model than the rule-enhanced LR-KGE model.}
	\label{tab_9}
	\begin{tabular}{|c|c|c|c|c|c|c|c|c|}
		\hline
		\textbf{Method}                                                                    & \multicolumn{2}{c|}{\textbf{LR-KGE}}                 & \multicolumn{2}{c|}{\textbf{AR-KGAT}}                & \multicolumn{2}{c|}{\textbf{LR-KGE}}                 & \multicolumn{2}{c|}{\textbf{AR-KGAT}}                \\ \hline
		\multirow{2}{*}{\textbf{Rule Category}}                                            & \multicolumn{4}{c|}{\textbf{Predicting head entities}}                                                      & \multicolumn{4}{c|}{\textbf{Predicting tail entities}}                                                      \\ \cline{2-9} 
		& \textbf{MRR} & \textbf{$\underline{\text{Hits}@10}$} & \textbf{MRR} & \textbf{$\underline{\text{Hits}@10}$} & \textbf{MRR} & \textbf{$\underline{\text{Hits}@10}$} & \textbf{MRR} & \textbf{$\underline{\text{Hits}@10}$} \\ \hline
		\textbf{none}                                                                      & 0.147        & 0.270                                 & 0.211        & 0.284                                 & 0.155        & 0.189                                 & 0.202        & 0.237                                 \\ \hline
		\textbf{\begin{tabular}[c]{@{}c@{}}one-to-one\\ (inference rule)\end{tabular}}     & 0.425        & 0.515                                 & 0.491        & 0.570                                 & 0.258        & 0.484                                 & 0.314        & 0.428                                 \\ \hline
		\textbf{\begin{tabular}[c]{@{}c@{}}one-to-one\\ (anti-symmetry rule)\end{tabular}} & 0.172        & 0.215                                 & 0.228        & 0.258                                 & 0.158        & 0.282                                 & 0.217        & 0.298                                 \\ \hline
		\textbf{n-to-one (n=2)}                                                            & 0.242        & 0.474                                 & 0.343        & 0.466                                 & 0.398        & 0.494                                 & 0.483        & 0.597                                 \\ \hline
		\textbf{overall}                                                                   & 0.388        & 0.554                                 & 0.501        & 0.562                                 & 0.313        & 0.505                                 & 0.437        & 0.483                                 \\ \hline
	\end{tabular}
\end{table*}

\begin{table*}
	\centering
	\caption{Effect of association rules on FB15k-237 and WN18RR for link prediction.}
	\label{tab_7}
	\begin{tabular}{|c|c|c|c|c|c|c|c|c|}
		\hline
		\multirow{2}{*}{\textbf{Model}} & \multicolumn{4}{c|}{\textbf{FB15K-237}}                                                                                            & \multicolumn{4}{c|}{\textbf{WN18RR}}                                                                                               \\ \cline{2-9} 
		& \textbf{MRR} & \textbf{$\underline{\text{Hits}@1}$} & \textbf{$\underline{\text{Hits}@3}$} & \textbf{$\underline{\text{Hits}@10}$} & \textbf{MRR} & \textbf{$\underline{\text{Hits}@1}$} & \textbf{$\underline{\text{Hits}@3}$} & \textbf{$\underline{\text{Hits}@10}$} \\ \hline
		\textbf{AR-KGAT-TriOnly}         &0.331	&0.244	&0.377	&0.497	&0.425	&0.364	&0.448	&0.535                            \\ \hline
		\textbf{AR-KGAT-OptOnly}        &0.355	&0.273	&0.395	&0.519	&0.456	&0.392	&0.446	&0.517                                 \\ \hline
		\textbf{AR-KGAT-AggOnly}        &0.394	&0.302	&0.446	&0.566	&0.484	&0.427	&0.487	&0.582                                \\ \hline
		\textbf{AR-KGAT}                    &0.442	&0.361	&0.483	&0.581	&0.518	&0.465	&0.540	&0.626                                 \\ \hline
	\end{tabular}
\end{table*}
\begin{table}
	\centering
	\caption{Effect of association rules on FB15k-237 and WN18RR for triple classification.}
	\label{tab_8}
	\begin{tabular}{|c|c|c|}
		\hline
		\multirow{2}{*}{\textbf{Model}} & \textbf{FB15K-237}               & \textbf{WN18RR}            \\ \cline{2-3} 
		& \textbf{MAP} & \textbf{MAP} \\ \hline
		\textbf{AR-KGAT-TriOnly}         &0.9091	&0.9523                            \\ \hline
		\textbf{AR-KGAT-OptOnly}        &0.9011	&0.9779                                 \\ \hline
		\textbf{AR-KGAT-AggOnly}        &0.9192	&0.9933                                \\ \hline
		\textbf{AR-KGAT}                    &0.9254	&0.9926                                \\ \hline
	\end{tabular}
\end{table}

Tables \ref{tab_7} and \ref{tab_8} show the experimental results on the test sets. The results indicate that both AR-KGAT-OptOnly and AR-KGAT-AggOnly outperform AR-KGAT-TriOnly by significant margins, implying the superiority of incorporating association logical rules. Compared to other variants, AR-KGAT-TriOnly shows the relatively poor result (both in WN18RR and FB15-237), denoting that the association rule mechanism is the critical element of the proposed model. In AR-KGAT-TriOnly, the representation relatedness is not explicitly modeled on the granularity of logical rules. A shallow model could be more suitable for the WN18RR dataset since it has only one type of rule. Thus, the performance of the proposed model can be lower for this dataset. Moreover, AR-KGAT-AggOnly always outperforms AR-KGAT-OptOnly, showing the capability of the graph attention network scenario to learn more predictive embeddings. In specific, the incorporation of the logical rules into the training of the aggregator is crucial to generate more accurate representations for entities, instead of only relying on a neural network to learn from the data. The proposed AR-KGAT combines the graph attention networks and the logical rules, thus providing outperforming performance over all the compared models. As shown in Table 7, the proposed AR-KGAT provides the performance improvement of the MRR by 23.3\%, and the HITS@10 by 28.2\% on the FB15k-237 dataset. Also, the proposed AR-KGAT shows a relatively non-obvious improvement in the WN18RR dataset: MRR and HITS@10 increase by 3.1\% and 0.3\%, respectively. The results show that more predictive embeddings are learned through the joint embedding of triplets and rules, especially for the dataset that contains multiple types of logical rules.

\subsubsection{Evaluation on Different Type of Association Rules}
In this section, the performance of the proposed model is analyzed for different types of rules. In this analysis, the FB15k-237 dataset is employed due to its diversity of logical rules set. As described in Section 4, the logical rules are classified into three types: one-to-one inference rules, one-to-one anti-symmetry rules, and n-to-one transitivity rules (n=2). We chose a logical rule powered knowledge graph embedding method, LR-KGE, for comparison. 

The results shown in Table \ref{tab_9} can be used to evaluate the performance and describe the behavior of different rules. As defined in Section 4, we divide rules into three types, for which the promotion percentage in FB15K-237 is 34.2\%, 7.9\%, and 18.0\% respectively over the variant with only triplets considered, demonstrating the superiority of incorporating logical rules. Moreover, it is shown that the proposed model is effective due to the modeling of multiple association rules. This also demonstrates that integrating rules into GNN allows the model to capture more complex knowledge information. However, the performance improvement is not always guaranteed, and the performance fluctuates since some association rules may contain noisy or conflict information with existing ones. From the results shown in Table \ref{tab_9}, we can observe that the one-to-one inference rule can get the better performance than other individual rules in predicting head entities task because this type of association rule contains more important correlations which promote the prediction performance. Similar to the previous results, in predicting tail entities, experimental results demonstrate that the two-to-one transitivity rule can get superior performance over other individual rules. By integrating all three types of logical rules, AR-KGAT achieves significant improvement across 4 out of 5 metrics compared to other variants. It also indicates that more ability to capture complicated correlations can be obtained by leveraging more information on multi-type logical rules, which is essential to learn more precise representations for entities and relations on knowledge graphs. The results also show that the performance can be improved by facilitating the alignment of different views in the proposed collaborative framework.  

\subsubsection{The analysis of the convergence}
\begin{figure}[h]
	\centering 
	\begin{minipage}{0.485\columnwidth}
		\includegraphics[width=\columnwidth]{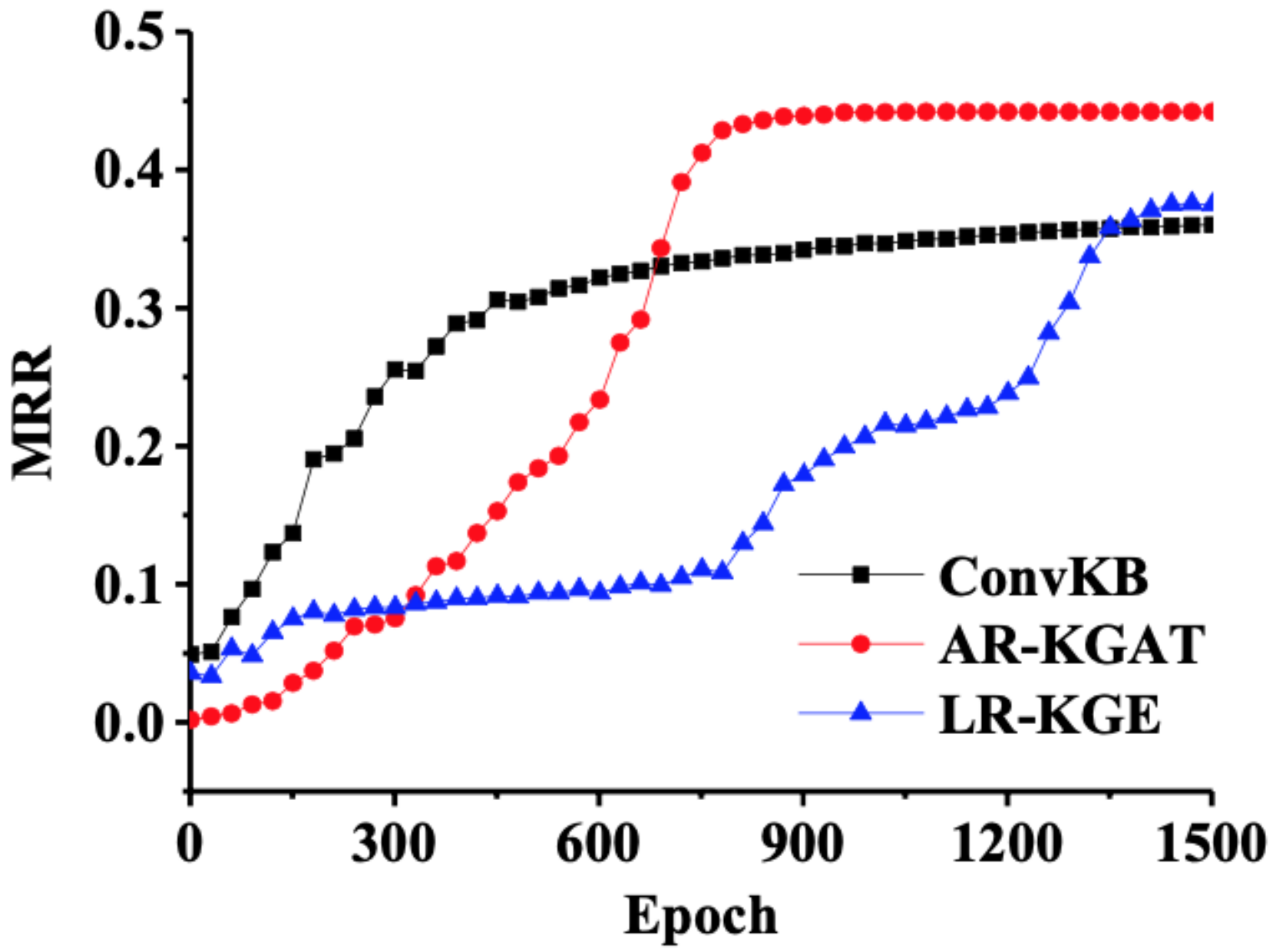}
		\centering 
		\subfigure{(a)\ MRR in FB-237}
	\end{minipage}
	\begin{minipage}{0.485\columnwidth}
		\includegraphics[width=0.9\columnwidth]{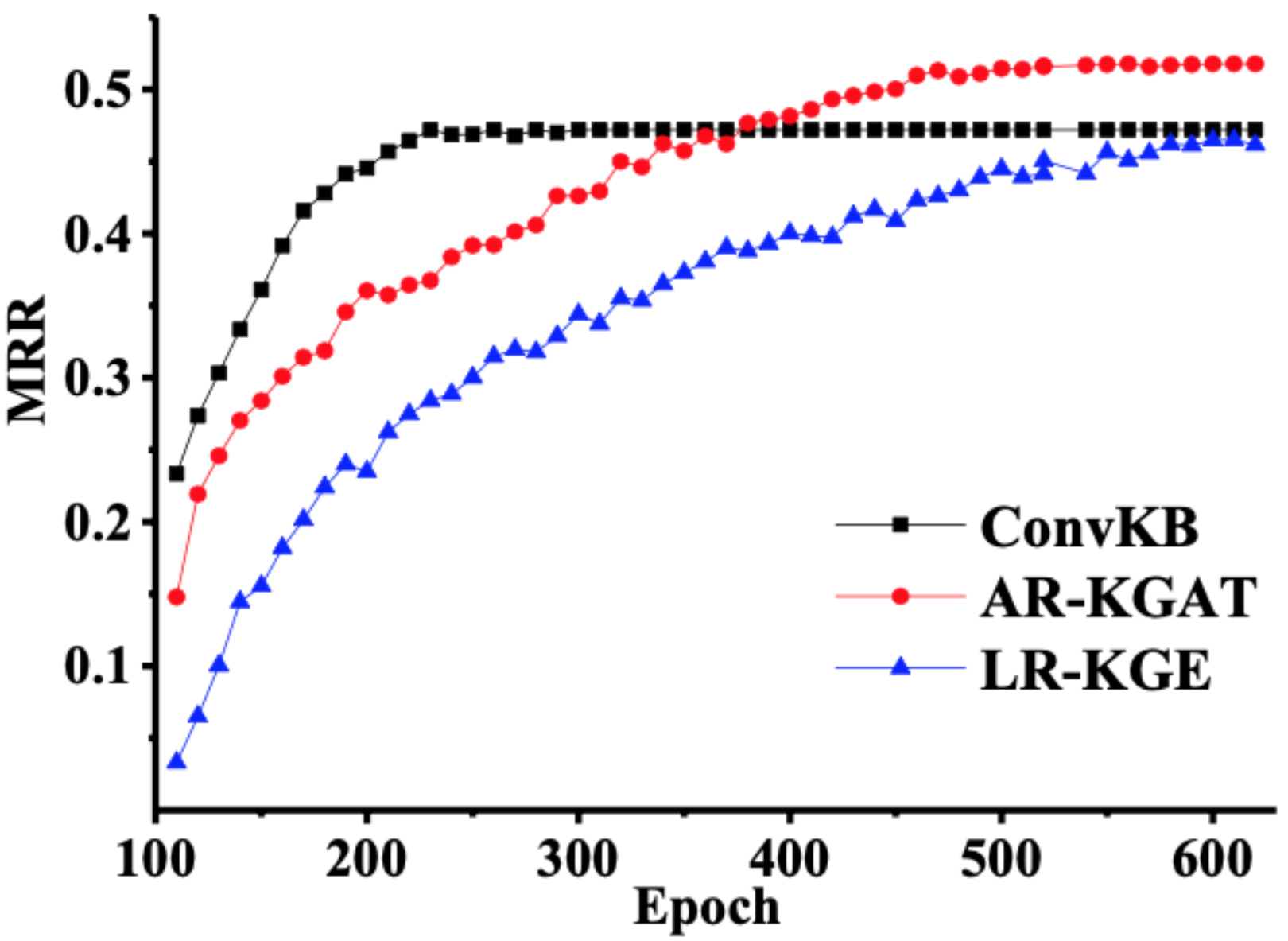}
		\centering 
		\subfigure{(b)\ MRR in WN18RR}
	\end{minipage}
	\caption{The convergence study of the three models in FB15K-237 and WN18RR datasets using the validation set. Due to the page limitation, only the results of MRR are reported here.}
	\label{fig_6}
\end{figure}

As shown in Figure \ref{fig_6}, in FB15K-237, our proposed AR-KGAT (the red line) provides better performance over ConvKB (the green line) and LR-KGE (the yellow line) after several epochs. Specifically, it can be seen that our proposed AR-KGAT shows continuously increasing performance until around 800 epochs. In contrast, the performances of the LR-KGE and ConvKB have converged around 1200 and 1,250 epochs in FB15K-237 respectively. However, in WN18RR, which has only one single type of rules, ConvKB achieves the fast coverage speed. The performances of ConvKB have converged about 250 epochs in WN18RR, and our model until nearly 500 epochs. Compared with relatively simple CNN based model, in our AR-KGAT model, not only a lot of significant parameters of graph attention network, but also a number of hyper-parameters of different types of association rules are required to be accurately refined according to lager epochs. Moreover, though different trends were observed with the FB15k-237 and WN18RR datasets, the metrics of our proposed AR-KGAT outperforms the other two models finally. The gap between these three models also shows that graph attention network and logical rules mechanism are two significant components in the proposed model. 

\subsection{The Analysis of Parameters Sensitivity}
The sensitivity of the AR-KGAT framework to parameters was analyzed, involving (1) number of neighbor relations, (2) model depth, (3) dimension of the learned node vectors, and (4) the proportion of unseen entities. In the following experiments, the different parameters are analyzed in terms of embedding performance.

\subsubsection{Number of Neighbor relations}
In this subsection, the effect of the number of neighbor relations to the performance is evaluated on three datasets. The node with a larger number means that it can receive more information from neighboring relations than other nodes with a smaller number. Specifically, the relation prediction task is analyzed with the varied number of neighbors from 3 to 10.
\begin{figure}[h]
	\centering 
	\begin{minipage}{0.485\columnwidth}
		\includegraphics[width=\columnwidth]{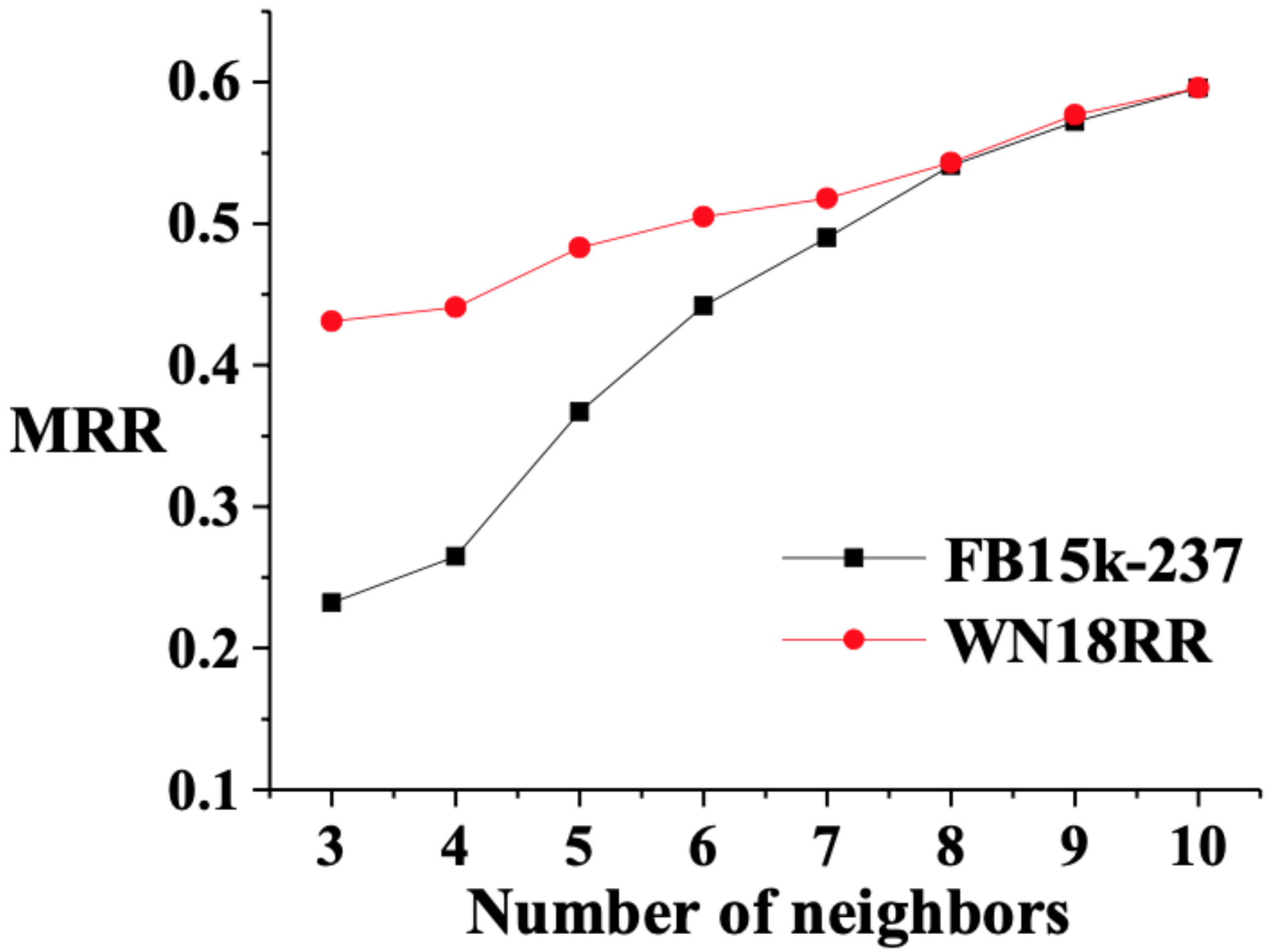}
	\end{minipage}
	\begin{minipage}{0.485\columnwidth}
		\includegraphics[width=\columnwidth]{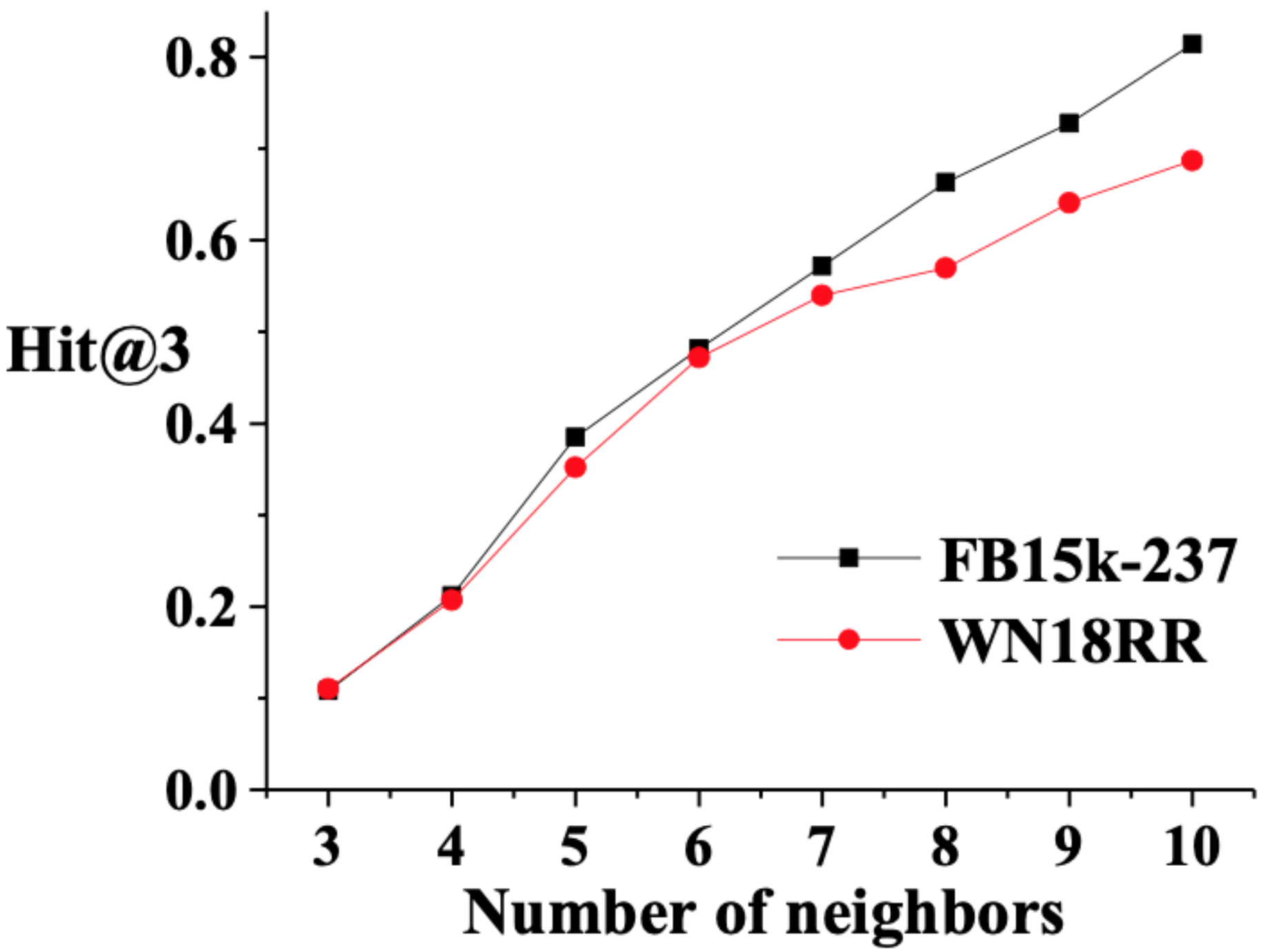}
	\end{minipage}
	
	\begin{minipage}{0.485\columnwidth}
		\includegraphics[width=\columnwidth]{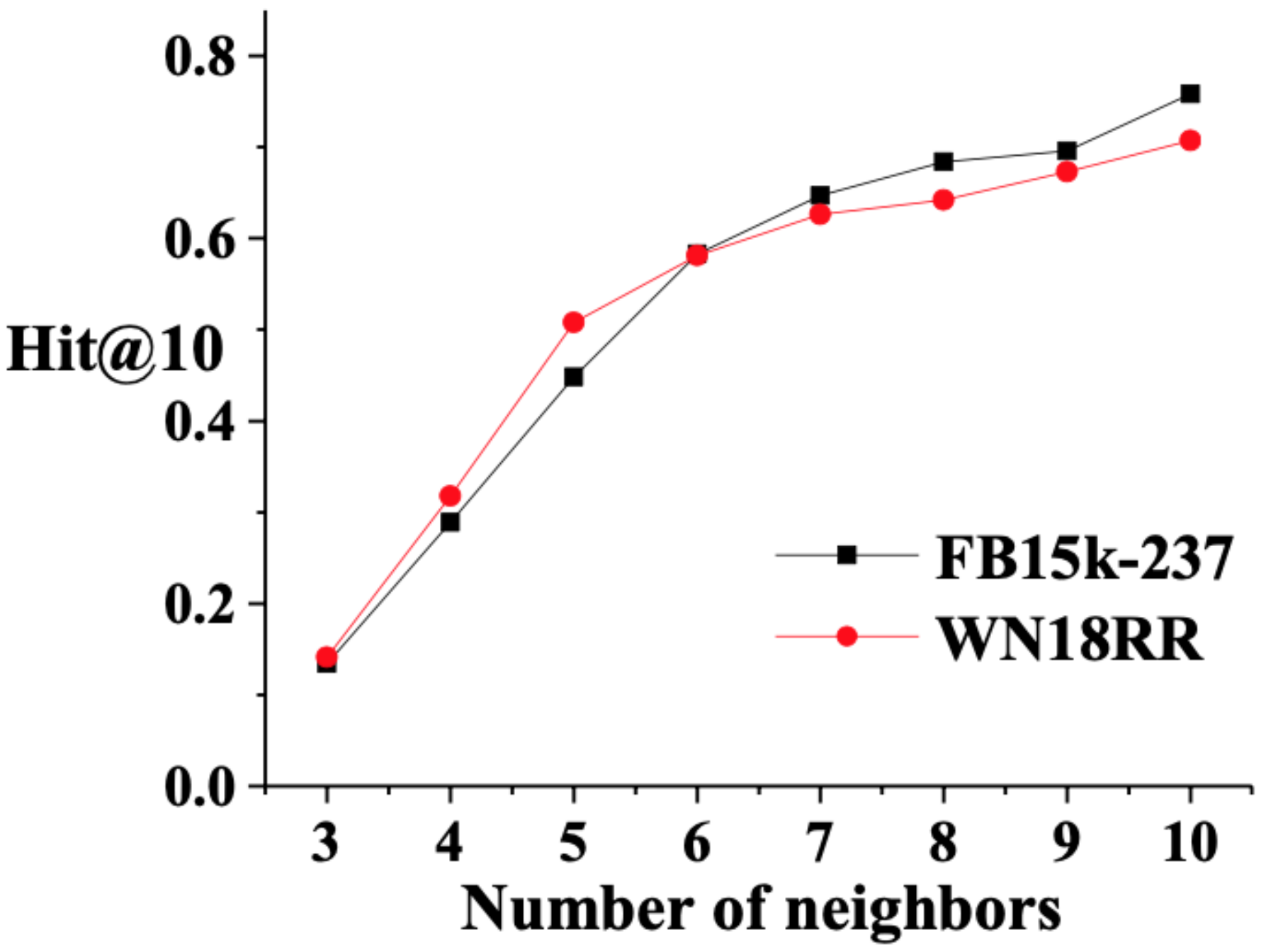}
	\end{minipage}
	\begin{minipage}{0.485\columnwidth}
		\includegraphics[width=\columnwidth]{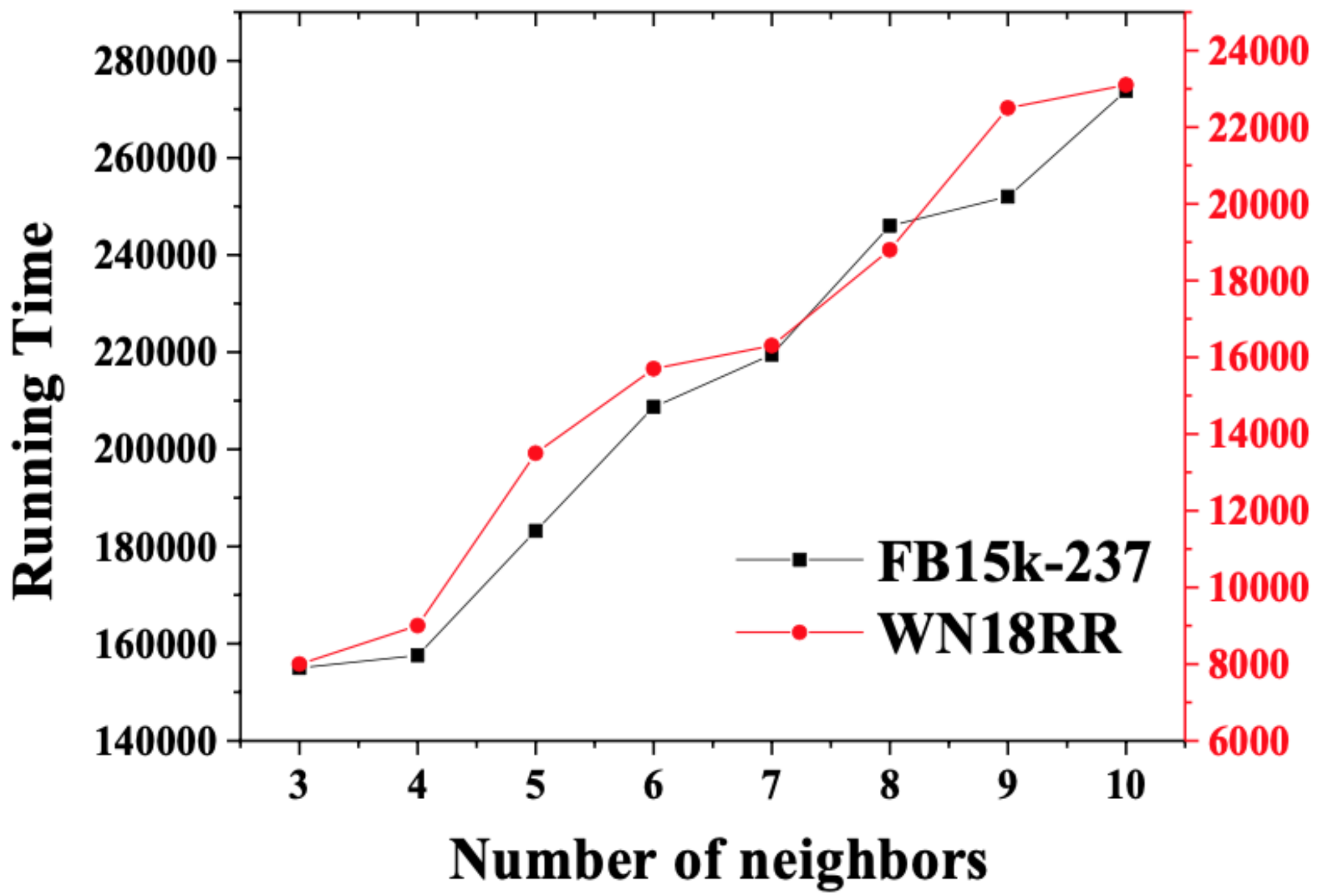}
	\end{minipage}
	\caption{Number of neighbor relations study using FB15k-237 and WN18RR datasets.}
	\label{fig_7}
\end{figure}

As shown in Figure \ref{fig_7}, different sets of nodes analyzed with a different number of neighbor relations. The average MRR, Hits@10, and Hits@3 scores are used in the analysis. For the different number of neighbor relations, the red line represents the performance for WN18RR, and the black line indicates the values for FB15k-237. The performance of the proposed model is increased according to the increased number of neighbors in terms of all the metrics, MRR, Hits@10, and Hits@3, showing that the information of neighbors can effectively enhance the representations of nodes. It is also shown that the personalized features of the entities can be modeled more effectively by using more neighbor information. The performance improvement, according to the increasing neighbors, tends to be steady for WN18RR. However, such an increment of the number of neighbors leads to the increment of the computational complexity, i.e., they are the trade-off. Therefore, to prevent impractical computation complexity due to the limited computational resources, the maximum numbers of neighbors are set as 8 for the FB15k-237 dataset and 6 for the WN18RR dataset.

\subsubsection{Effect of Model Depth}
We vary the depth of KGAT (\textit{e.g., L}) to investigate the efficiency of usage of multiple embedding propagation layers. In particular, the layer number is searched in the range of \{1, 2, 3\}. We use KGAT-1 to indicate the model with one layer and similar notations for others. We summarize the results in Table 10 and have the following observations: 
\begin{table*}
	\centering
	\caption{Effect of association rules on FB15k-237 and WN18RR for link prediction.}
	\label{tab_10}
	\begin{tabular}{|c|c|c|c|c|c|c|c|c|}
		\hline
		\multirow{2}{*}{\textbf{Model}} & \multicolumn{4}{c|}{\textbf{FB15K-237}}                                                                                            & \multicolumn{4}{c|}{\textbf{WN18RR}}                                                                                               \\ \cline{2-9} 
		& \textbf{MRR} & \textbf{$\underline{\text{Hits}@1}$} & \textbf{$\underline{\text{Hits}@3}$} & \textbf{$\underline{\text{Hits}@10}$} & \textbf{MRR} & \textbf{$\underline{\text{Hits}@1}$} & \textbf{$\underline{\text{Hits}@3}$} & \textbf{$\underline{\text{Hits}@10}$} \\ \hline
		\textbf{AR-KGAT-1}         &0.370	&0.303	&0.412	&0.517	&0.467	&0.419	&0.492	&0.579          \\ \hline
		\textbf{AR-KGAT-2}        &0.442	&0.361	&0.483	&0.581	&0.518	&0.465	&0.550	&0.626           \\ \hline
		\textbf{AR-KGAT-3}        &0.448	&0.361	&0.497	&0.602	&0.527	&0.493	&0.568	&0.630           \\ \hline
	\end{tabular}
\end{table*}

Increasing the depth of KGAT is capable of boosting the performance substantially. Clearly, KGAT-2 achieves consistent improvement over KGAT-1 across all the board. The results demonstrate that appropriate depth is essential for training graph attention networks to characterize rich high-order semantic relations between entities. We attribute the improvements to the effective modeling of high-order relation between entities, carried by the first- and second-order connectivities, respectively. Further stacking one more layer over KGAT-2, we observe that KGAT-3 only achieve marginal improvements. It suggests that taking into account first-order and second-order relations among entities simultaneously could be sufficient to the demands of most downstream applications. 

\subsubsection{Effect of Embedding Dimension}
\begin{figure}[h]
	\centering 
	\begin{minipage}{0.485\columnwidth}
		\includegraphics[width=\columnwidth]{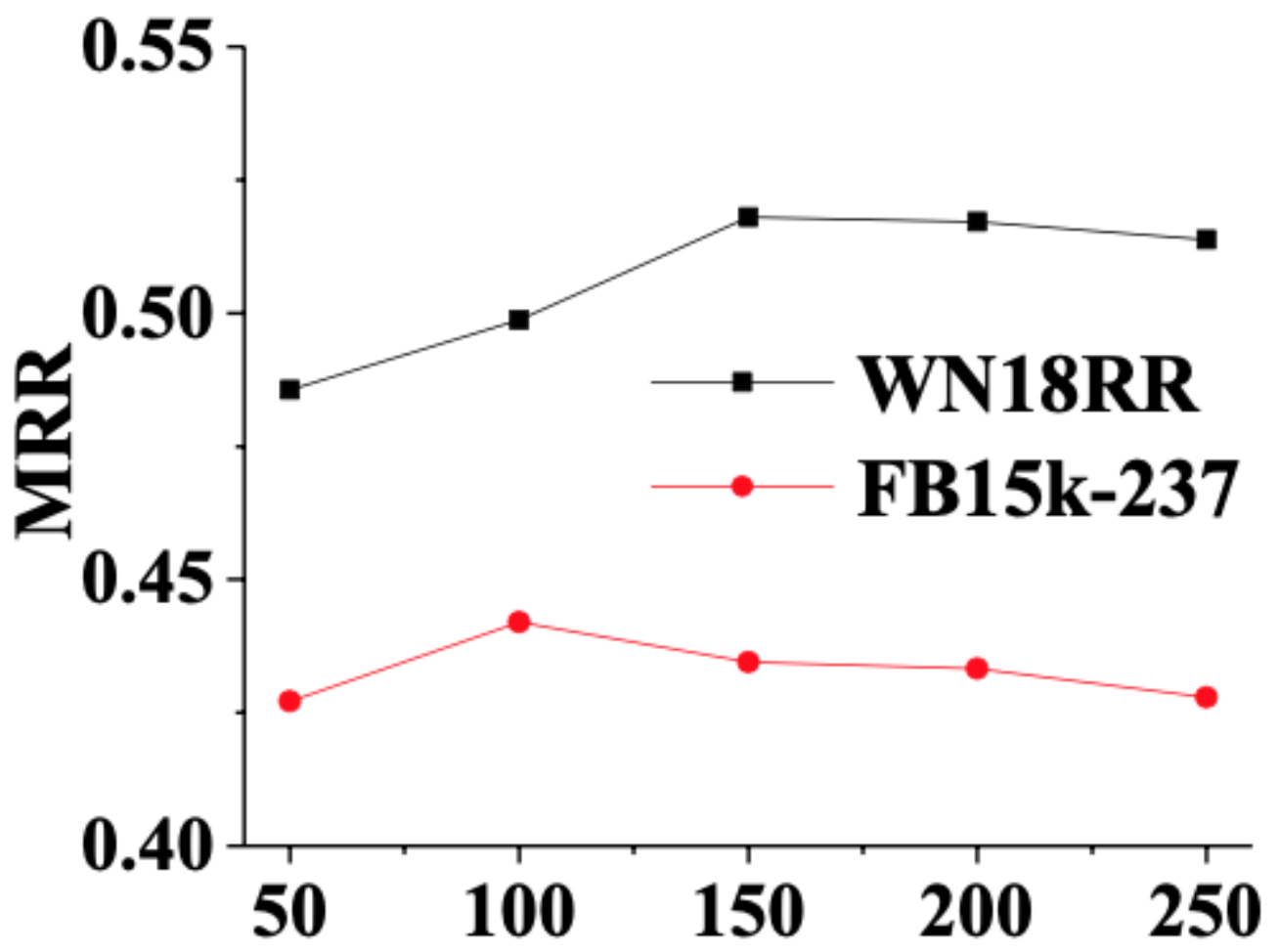}
	\end{minipage}
	\begin{minipage}{0.485\columnwidth}
		\includegraphics[width=1.05\columnwidth]{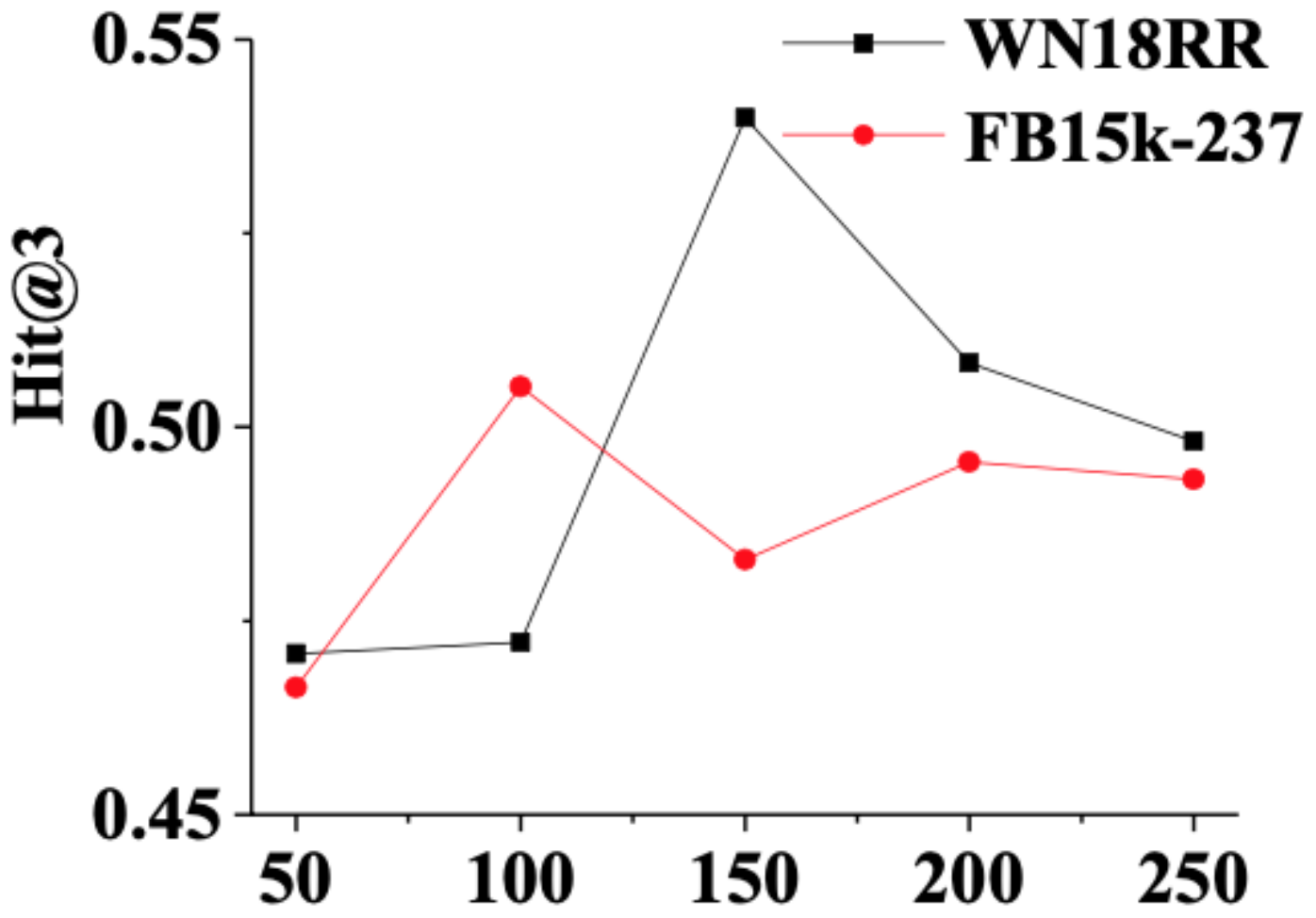}
	\end{minipage}
	
	\begin{minipage}{0.485\columnwidth}
		\includegraphics[width=\columnwidth]{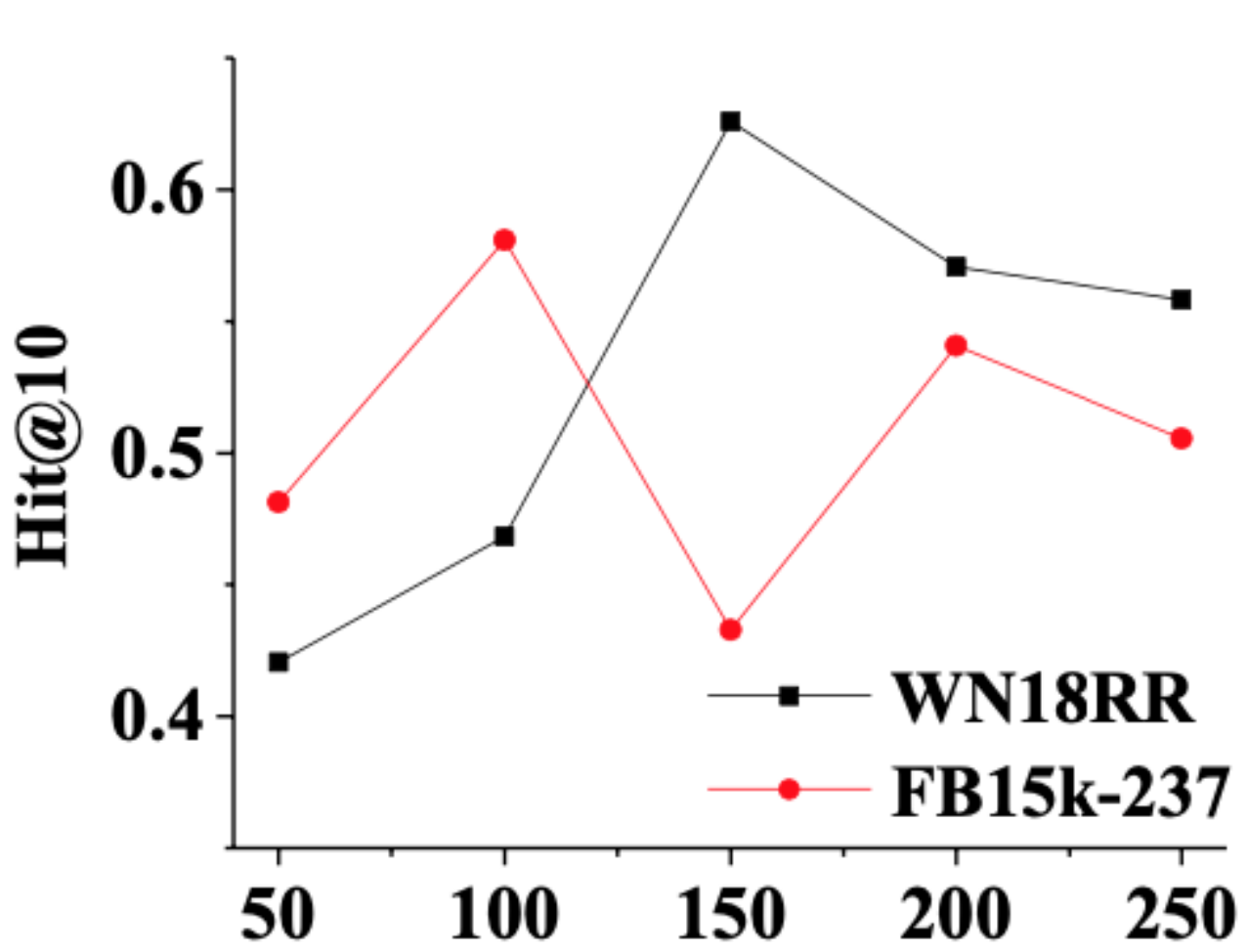}
	\end{minipage}
	\caption{Effect of the varying embedding dimension ($d$).}
	\label{fig_8}
\end{figure}

Relation prediction performance was analyzed when entity embedding dimension (denoted by d and learned by AR-KGAT) varies on both datasets of FB15k-237 and WN18RR. The overall results are summarized in Figure \ref{fig_8}. 

We observe that on increasing the dimension of embeddings, the value of MRR and Hit@k grows first. Such a phenomenon occurs because this model requires a proper dimension to preserve useful semantic information. However, it then remains unchanged and degrades as the dimension of entity representation is increased beyond a certain limit. We hypothesize that this is due to the over parameterization of the model. Moreover, if the dimension is excessively large, noisy information may be added, which consequently leads to worse performances and brings extra storage burden. Based on the experimental findings above, our proposed AR-KGAT needs a proper dimension to encode rich semantic information, and too large a dimension may introduce additional redundancies. Hence, we set the dimension of embeddings as 150 on WN18RR, and 100 on FB15k-237. 

\subsubsection{Influence of the proportion of unseen entities}
When the ratio of the unseen entities over the training entities increases (i.e., the observed knowledge graph becomes sparser), the performance of the model generally deteriorated. The effect of the sparse knowledge graph in the proposed model is analyzed in the link prediction task for different sample rates of datasets. 
\begin{figure}[h]
	\centering 
	\begin{minipage}{0.485\columnwidth}
		\includegraphics[width=\columnwidth]{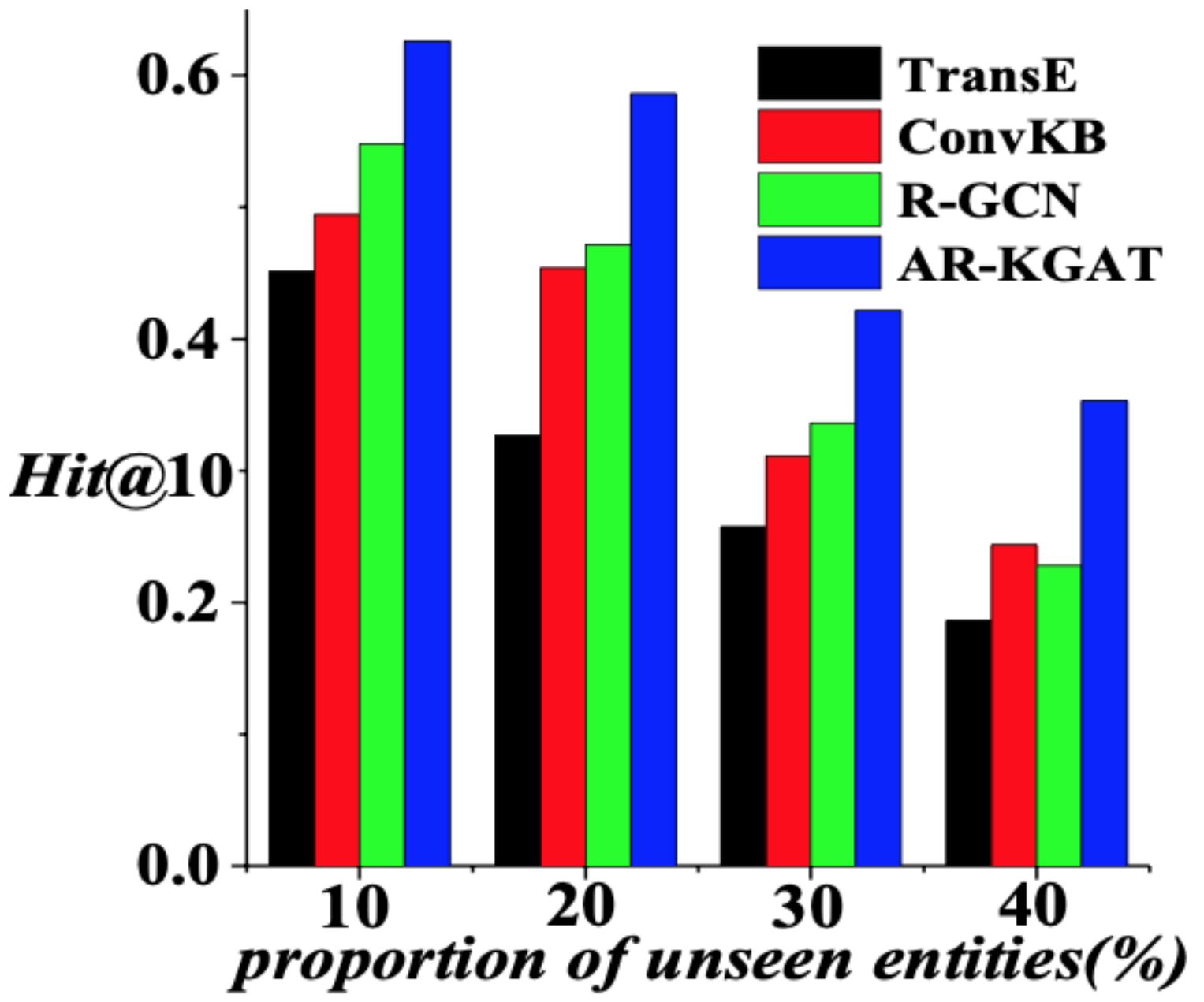}
	\end{minipage}
	\begin{minipage}{0.485\columnwidth}
		\includegraphics[width=\columnwidth]{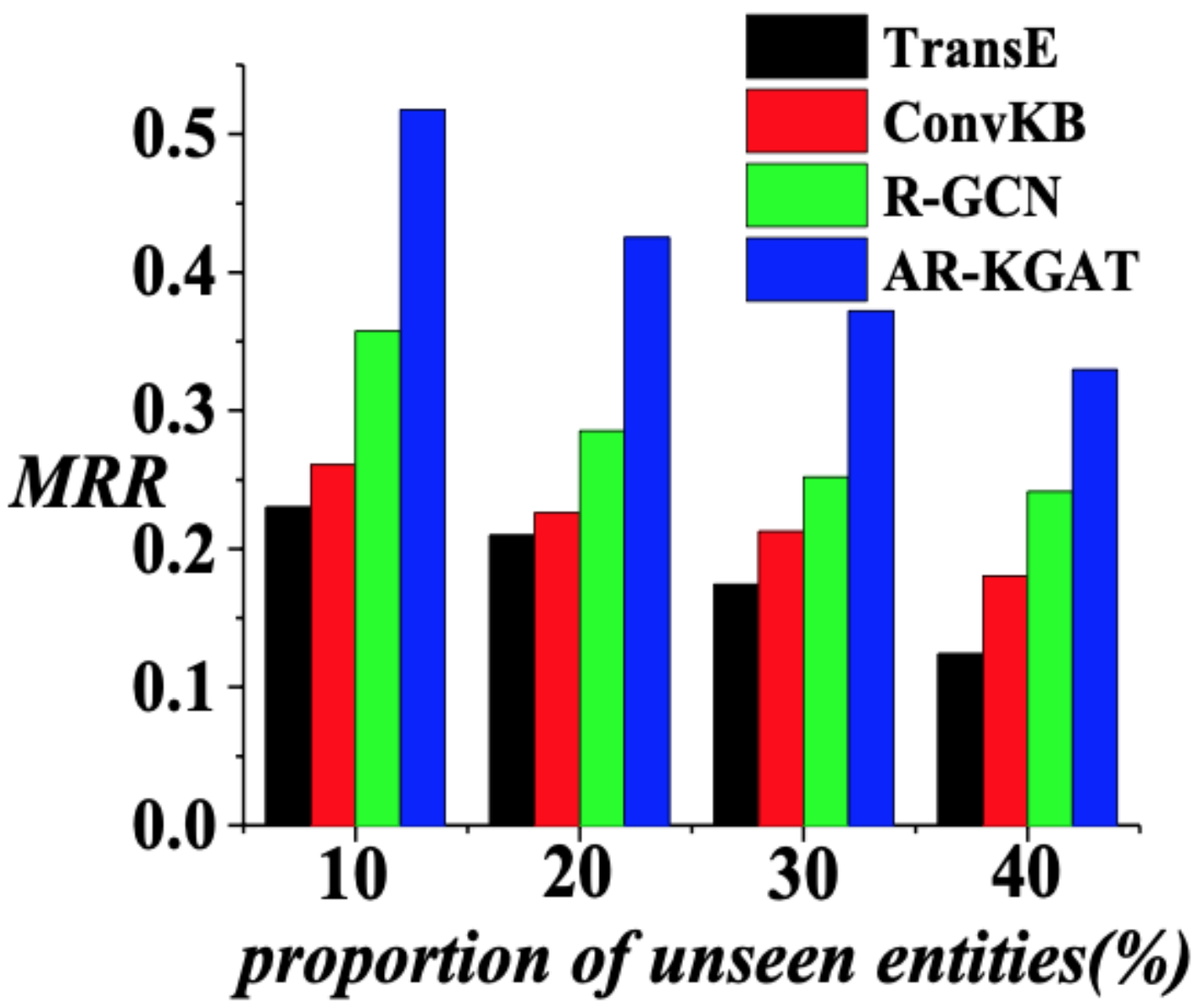}
	\end{minipage}
	\caption{Effect of the proportion of unseen entities on FB15k-237 dataset.}
	\label{fig_9}
\end{figure}

The results are displayed in Figure \ref{fig_9}. We observe that the increasing proportion of unseen entities certainly has a negative impact on all models. However, the performance of our model does not decrease as drastically as that of R-GCN, TransE and ConvKB (three SOTA baselines), indicating that AR-KGAT is more robust on sparse KGs. 

\section{Conclusion and future work}
\label{section7}

This paper proposes a joint embedding framework, AR-KGAT, for knowledge graphs and logical rules. The key idea is to integrate triplets and association rules in the knowledge graph attention network framework to generate effective representations. Specifically, the graph attention mechanisms are generalized and extended so that both entity and relation features are captured in a multi-hop neighborhood of a given entity. In our proposed aggregator, the weights of a coarse relation level and a fine neighbor level are estimated by logical rules and neural attention networks, respectively. Moreover, the embedding task is then conducted by minimizing the loss function on both complex and atomic formulas. In this way, the learned embeddings are certainly more useful for knowledge acquisition and inference, which are compatible with triplets and rules. The proposed AR-KGAT is evaluated for two datasets. Each dataset is used for both the triplet classification and link prediction tasks. Experimental results indicate that significant and consistent performance improvement is achieved through the joint embedding over state-of-the-art models on two typical KG completion tasks. The detailed and exhaustive empirical analysis gives insight into the superiority of the proposed method for relation prediction on KGs. Future works will include extending the proposed method to hierarchical graphs, capturing higher-order relations between entities in our graph attention model. Further, more types of effective rules that exist in data will be investigated. The embedding of multiple-source knowledge graphs will also be explored, e.g., jointly embedding Freebase and YAGO.

\section*{Acknowledgments}

	The work was supported by the National Natural Science Foundation of China [grant numbers: 61876138, 61602354]. Any opinions, findings and conclusions expressed here are those of the authors and do not necessarily reflect the views of the funding agencies.

\bibliographystyle{IEEEtran}
\bibliography{refs}

\end{document}